\documentclass[journal]{IEEEtran}
\usepackage{epsfig,amsfonts,amsbsy,bm,mathrsfs}
\usepackage{mathrsfs} 
\usepackage{graphicx,cite,amssymb,amsmath,bm}
\usepackage{mathtools}
\usepackage{amsmath} 
\usepackage{etex}
\usepackage{amsmath,amssymb}
\usepackage{lipsum}
\usepackage{array,verbatim}
\usepackage{makecell}
\usepackage{siunitx}
\usepackage{float}
\usepackage{xcolor}
\usepackage{colortbl}
\usepackage{tikz}
\usepackage{ctable}
\usepackage[english]{babel}
\usepackage{color}
\usepackage[utf8]{inputenc}
\usepackage[T1]{fontenc}
\usepackage{balance}
\usepackage{balance}
\usepackage{perpage}
\usepackage{algorithm}
\usepackage[noend]{algpseudocode}
\usepackage{subcaption}
\usepackage{graphbox,graphicx}
\usepackage{xcolor}
\usepackage{bbm}

\newtheorem{proposition}{Proposition}

\newtheorem{remark}{Remark}

\addto\captionsenglish{}
\makeatletter
\def\algbackskip{\hskip-\ALG@thistlm}
\makeatother

\newcommand{\mep}{{x}}
\newcommand{\row}{\mathsf{r}}
\newcommand{\col}{\mathsf{c}}

\newcommand{\Q}{\mathsf{Q}}
\newcommand{\G}{\mathsf{G}}

\renewcommand{\a}{\boldsymbol{a}}
\renewcommand{\b}{\boldsymbol{b}}

\newcommand{\nc}{n}

\newcommand{\kc}{k}

\newcommand{\lr}{\boldsymbol{r}}

\newcommand{\lc}{\boldsymbol{c}}

\newcommand{\rr}{\boldsymbol{R}}

\newcommand{\cc}{\boldsymbol{C}}

\newcommand{\dmin}{d_\mathsf{min}}

\newcommand{\lalone}{\boldsymbol{L}}

\newcommand{\BB}{\mathsf{B}}

\newcommand{\mut}{\tilde{\mu}}

\definecolor{mygreen}{rgb}{0.0, 0.45, 0.0}

\def\forcemath#1{\ifmmode #1 \else $#1$\fi}

\newcommand{\Pue}{P^{\mathsf{e}}}
\newcommand{\Que}{Q^{\mathsf{e}}}
\newcommand{\Puc}{P^{\mathsf{c}}}
\newcommand{\Quc}{Q^{\mathsf{c}}}

\newcommand{\Puep}{P^{\mathsf{\epsilon}}}
\newcommand{\Quep}{Q^{\mathsf{\epsilon}}}

\newcommand{\ham}{\mathsf{d}_\mathsf{H}}

\newcommand{\opt}{\tilde{l}}

\newcommand{\Eb}{\mathsf{E_b}}
\newcommand{\Es}{\mathsf{E_s}}
\newcommand{\No}{\mathsf{N_0}} 
\definecolor{armygreen}{rgb}{0.29, 0.33, 0.13}
\newcommand{\SH}{\textcolor{black}}
\newcommand{\SHm}{\textcolor{black}}
\newcommand{\GL}{\textcolor{black}}
\newcommand{\AG}{\textcolor{black}}
\newcommand{\AGc}{\textcolor{black}}

\ifCLASSINFOpdf
\else
\fi

\newcommand\myeqa{\stackrel{\mathclap{\normalfont\mbox{\scriptsize $(a)$}}}{=}}
\newcommand\myeqb{\stackrel{\mathclap{\normalfont\mbox{\scriptsize $(b)$}}}{=}}
\newcommand\myeqc{\stackrel{\mathclap{\normalfont\mbox{\scriptsize$(c)$}}}{=}}
\newcommand\myeqd{\stackrel{\mathclap{\normalfont\mbox{\scriptsize$(d)$}}}{=}}
\newcommand\myeqe{\stackrel{\mathclap{\normalfont\mbox{\scriptsize$(e)$}}}{=}}

\let\originalleft\left
\let\originalright\right
\renewcommand{\left}{\mathopen{}\mathclose\bgroup\originalleft}
\renewcommand{\right}{\aftergroup\egroup\originalright}

\newcommand\scalemath[2]{\scalebox{#1}{\mbox{\ensuremath{\displaystyle #2}}}}   

\begin{document}

\title{Refined Reliability Combining for Binary Message Passing Decoding of Product Codes}
\author{
	Alireza Sheikh \IEEEmembership{Member, IEEE}, Alexandre Graell i Amat, \IEEEmembership{Senior Member, IEEE}, Gianluigi~Liva,~\IEEEmembership{Senior~Member,~IEEE}, and Alex Alvarado \IEEEmembership{Senior Member, IEEE}. 
	\thanks{A. Sheikh and A. Alvarado are with the Department of Electrical Engineering, Eindhoven University of Technology, PO Box 513
		5600 MB Eindhoven, The Netherlands (emails: \{asheikh, a.alvarado\}@tue.nl). The work of A. Sheikh and A. Alvarado has received funding from the European Research Council (ERC) under the European Union's Horizon 2020 research and innovation programme (grant agreement No 757791).}	
	\thanks{A. Graell i Amat is with the Department of Electrical Engineering, Chalmers University of Technology, SE-41296 Gothenburg, Sweden (email: alexandre.graell@chalmers.se).}
	\thanks{G. Liva is with the Institute of Communications and
		Navigation of the German Aerospace Center (DLR), M\"unchner Strasse 20, 82234 We{\ss}ling, Germany (email: gianluigi.liva@dlr.de).}}

\maketitle

\begin{abstract}
	
We propose a novel soft-aided iterative decoding algorithm for product codes (PCs). The proposed algorithm, named iterative bounded distance decoding with combined reliability (iBDD-CR), enhances the conventional iterative bounded distance decoding (iBDD) of PCs by exploiting some level of soft information. In particular, iBDD-CR can be seen as a modification of iBDD where the hard decisions of the row and column decoders  are made based on \GL{a reliability estimate} of the BDD outputs. \GL{The reliability estimates} are derived using extrinsic message passing for generalized low-density-parity check (GLDPC) ensembles, \AG{which  encompass PCs}. 
We perform a density evolution analysis of iBDD-CR  for transmission over the additive white Gaussian noise channel for the GLDPC ensemble. We consider both binary transmission and bit-interleaved coded modulation with quadrature amplitude modulation. We show that iBDD-CR achieves performance gains up to $0.51$ dB compared to iBDD with the same internal decoder data flow.
This makes the algorithm an attractive solution for very high-throughput applications such as fiber-optic communications.  
 

\end{abstract}


\section{Introduction}

The rediscovery of soft-decision (SD) iterative decoding algorithms and graph-based codes in the early 1990s allowed for the first time performance close to the theoretical limits with relatively low complexity. Turbo \GL{\cite{Berrouetal93}}, and low-density parity-check (LDPC) \GL{\cite{Gallager63:LDPC}} codes were soon widely adopted in communications standards. 
However, the ever more demanding requirements in terms  of  throughput and  power consumption, asks for new coding solutions.
For example, current optical coherent transceivers operate at data rates of 400 Gbps and the next frontier  is $1$~Tbps. Scaling conventional SD iterative decoders to such high throughputs  is very challenging.  This has spurred a great deal of research in the last few years on designing low-complexity coding and decoding schemes that can operate at very high throughputs while yet achieving performance close to that of conventional SD iterative schemes \cite{Cushon2016,Ste19}. 

A key observation is that the main limitation to achieve very high throughputs arises from the high internal decoder data flow of SD decoders \cite{staircase_frank}. For this reason hard-decision (HD) decoders are appealing solutions when  high throughputs are sought, as the internal decoder data flow can be kept reasonably small. For example, HD coding for optical communications is usually
based on product-like codes with high rate  Bose-Chaudhuri-Hocquenghem (BCH) component codes, which can be efficiently decoded via bounded distance
decoding (BDD) \cite{staircase_frank,JianPfister2017}. The decoding is then performed based on BDD of the component
codes and iterating between the row and column decoders, which we refer here to as iterative BDD (iBDD). The very high throughputs of HD product-like coding schemes, however, are achieved at the expense of a significant performance loss compared to their LDPC code counterparts, decoded via SD belief propagation. 

Recently, several works have focused on the improvement of the performance of conventional iBDD of product-like codes  targeting specifically very high throughputs \cite{Hag18,sheikhTCOM19,She18b,She19,YibitTCOM}. The underlying idea in all these works is to generate some level of reliability information \GL{within} the decoder to assist the conventional BDD. The result is an enhanced performance compared to iBDD while keeping the internal data flow low. Among the algorithms in  \cite{Hag18,sheikhTCOM19,YibitTCOM}, iBDD with scaled reliability (iBDD-SR) \cite{sheikhTCOM19} is the one yielding the smallest increase in complexity, yet achieving about $0.3$ dB performance improvement compared to iBDD for binary transmission. 
Following the principle introduced in \cite{Lech12}, iBDD-SR is based on binary message passing (BMP) between component decoders and \AG{generates reliability information at the BDD output by scaling the decisions according to a reliability estimate of the decision. The reliability information, in the form of log-likelihood ratios is then added} to the corresponding channel LLRs to form refined bit estimates (see \cite[Fig.~2]{sheikhTCOM19}). The decoding schemes in \cite{She18b,She19,YibitTCOM,GabrieleSABMSR2019}, on the other hand, yield some additional coding gains but require the knowledge of the least reliable bits in the decoding process, and hence entail further complexity. \GL{Alternative constructions for high-throughput applications include coarsely-quantized low-density parity-check decoders \cite{Lech12,Ben18,Ste19}, two-stage decoders \cite{Montorsi2018}, and SD-HD hybrid schemes based on concatenating a relatively weak SD code and an outer HD product-like code \cite{Ksch17,Ksch18}.}

This paper extends our previous work \cite{sheikhTCOM19} in three different directions: 
\GL{\begin{itemize}
\item[i.] We \GL{derive a more accurate estimate of} the reliability of the BDD outputs, which allows us to derive \GL{an improved} combining \GL{rule for} the BDD outputs and the channel LLRs. The resulting decoding algorithm, dubbed iterative bounded distance decoding with combined reliability (iBDD-CR) is shown to outperform iBDD-SR and, interestingly, also an idealized (genie-aided) iBDD that prevents miscorrections. The \GL{combining rule} can be implemented by means of a small lookup table (LUT).
\item[ii.] We ad\GL{a}pt the algorithm to bit interleaved coded modulation (BICM) with \SH{nonbinary} modulation. For both binary and BICM transmission, we derive the density evolution (DE) equations for iBDD-CR, which provides an amenable analysis of the algorithm. 
\item[iii.] We \SH{evaluate} the effect of quantizing the channel \GL{LLRs} and show that iBDD-CR has low sensitivity to quantization.
\end{itemize}}
\GL{The combining rule derived in this work is based on DE analysis of (regular) generalized LDPC (GLDPC) codes ensembles that contain product codes \SH{(PCs)} as ensemble members, under extrinsic message passing decoding. For the ensembles under consideration, the derived rule is optimal in the limit of large block lengths. Remarkably, the proposed combining rule provides a substantial coding gain with respect to iBDD-SR also when applied to the decoding of \SH{PCs}, with gains that are consistent with the DE analysis findings.}
We show that 
iBDD-CR yields gains up to $0.51$ dB compared to iBDD with \GL{an} identical decoder data flow, 
making the algorithm particularly appealing for high-throughput applications.

\medskip

\noindent \emph{Notation:} We use boldface letters to denote vectors
and matrices, e.g., $\boldsymbol{x}$  and $\boldsymbol{X}=[x_{i,j}]$, with $x_{i,j}$ representing the element corresponding to the $i$-th row and $j$-th column of $\boldsymbol{X}$. 
Moreover, $\boldsymbol{X}_{i,:}$ denotes the $i$-th row of $\boldsymbol{X}$.   
$|a|$ denotes the absolute value of $a$,
$\left\lfloor a \right\rfloor$ the largest integer smaller than or equal to $a$, and $\left\lceil a \right\rceil$ the smallest integer larger than or equal to $a$. $(\cdot)^{\mathsf{T}}$ denotes the transpose operation. We also define $\hat l \in \{\pm1\}$ as the sign of value $l$.  $\mathbb{R}$ is the set of real numbers and $p(\cdot)$ is the probability distribution or probability mass function of the continuous or discrete random variables (RVs), respectively. 
A Gaussian distribution with mean $\mu$ and variance $\sigma^2$ is denoted by $\mathcal{N}(\mu ,\sigma^2)$. Furthermore, 
$\G (\lambda ;\mu ,{\sigma ^2}) \triangleq \frac{1}{{\sqrt {2\pi } \sigma }}\exp ( - \frac{{{{\left( {\lambda  - \mu } \right)}^2}}}{{2{\sigma ^2}}})$ stands for the Gaussian function with mean and variance $\mu$ and $\sigma^2$. We also denote by $\Q(x)\triangleq\frac{1}{\sqrt{2\pi}}\int_x^\infty \mathrm{e}^{\frac{-\xi^2}{2}}\mathrm{d}\xi$ the tail \AG{probability} of the standard Gaussian distribution. The Hamming distance between vectors $\a$ and $\b$ is denoted by $\ham(\a,\b)$. Finally, \GL{we define}
\begin{align*}
\bar{\mathbbm{U}}(x)=  \left\{ 
\begin{array}{ll}
1 & \text{if } x<0\\
0 & \text{otherwise}.
\end{array}
\right.
\end{align*}

\section{Preliminaries}
\label{sys_mod} 

We consider binary PCs with BCH component codes. Let $\mathcal{C}$ be an $(\nc,\kc)$ BCH code with minimum Hamming distance $\dmin$ built over the Galois field $\text{GF}(2^v)$ with (even) block length $\nc$ and information block length $\kc$ given by
\begin{align}
\nc=2^v-1, \;\;\; \kc=2^v-vt-1, \label{k_formula}
\end{align} 
where $t \buildrel \Delta \over =  \left\lfloor {\frac{{{\dmin} - 1}}{2}} \right\rfloor$ \AG{is} the error correcting capability of the code.

A (two-dimensional) PC with parameters $(\nc^2,\kc^2)$ and code rate  $R=\kc^2/\nc^2$ is defined as the set of all $\nc\times\nc$ arrays such that each row and each column in the array is a codeword of $\mathcal{C}$. Accordingly, a codeword can be defined as a binary matrix $\cc=[c_{i,j}]$.  
For \AG{ease of explanation, assume first} transmission over the binary input additive white Gaussian noise (bi-AWGN) channel \AG{(in Sec.~\ref{QAM_BICM} we also consider a bit-interleaved coded modulation (BICM))}. The output of the bi-AWGN channel corresponding to code bit $c_{i,j}$ is thus given by
\begin{align}\label{channel_inst}
y_{i,j}=x_{i,j}+n_{i,j},
\end{align}  
where $x_{i,j}=(-1)^{c_{i,j}}$ and $n_{i,j}\sim \mathcal{N}(0,\sigma^2)$. For a given $\Eb/\No$ and code rate $R$, the \AG{noise variance is $\sigma^2=(2 R
\Eb/\No)^{-1}$. The signal-to-noise ratio \SH{(SNR)} per symbol is  $\Es/\No=\frac{1}{2\sigma^2}$}. Let $\lalone=[l_{i,j}]$ be the matrix of channel \AG{LLRs} and $\rr=[r_{i,j}]$ the matrix of hard decisions at the output of the channel, i.e., $r_{i,j}$ is obtained taking the sign of $l_{i,j}$ and mapping $- 1 \mapsto 1$ and $+ 1 \mapsto 0$. We denote this mapping by $\BB(\cdot)$, i.e., $r_{i,j}=\BB(l_{i,j})$.  
With some abuse of notation we will write $\rr=\BB(\lalone)$.

PCs are conventionally decoded using \AG{BDD} of the component codes. Here we briefly explain BDD. Consider the decoding of an arbitrary row component, which is an $1 \times n$ array. \AG{Specifically}, assume decoding of the transmitted component codeword $\lc=(c_1,\ldots,c_{\nc})$ based on the hard-detected bits at the channel output,
$\lr=(r_1,\ldots,r_{\nc})$. BDD corrects all
error patterns with Hamming weight up to the error-correcting
capability of the code $t$. If the
weight of the error pattern is larger than $t$ and there exists
another codeword $\tilde{\lc} \in \mathcal{C}$ with
$\mathsf{d_H}(\tilde{\lc},\lr)\le t$, then BDD erroneously decodes $\lr$ onto
$\tilde{\lc}$ and a so-called \emph{miscorrection} \AG{occurs}.  Otherwise, if
such a codeword does not exist, BDD fails. \GL{\AG{In this case, the bounded distance (BD) decoder}  may output its input $\lr$, or it may declare a decoding failure (often referred to as an \emph{erasure}).}
The decoding of PCs can be accomplished in an iterative fashion based on  BDD of the component codes and iterating between the row and column decoders, \AG{which we refer to as iBDD}. \GL{In the case of iBDD, if a local \AG{BD decoder} fails, it outputs the input vector.}
 
\begin{figure*}[!t] \centering 
	\includegraphics[scale=0.9]{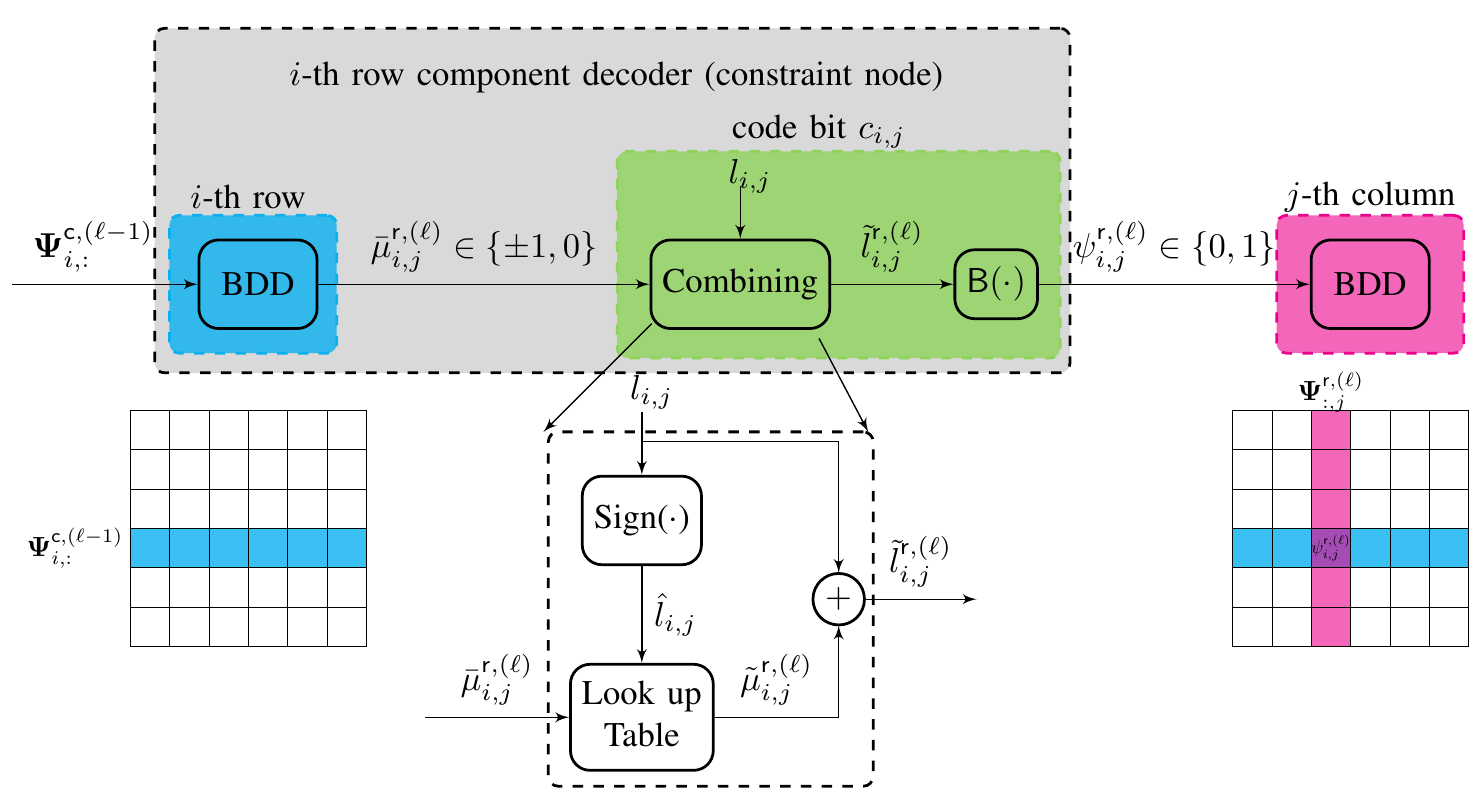}  
	\caption{Block diagram \GL{illustrating the} message passing in iBDD-CR corresponding to $i$th row decoding at iteration $\ell$. The diagram shows the decision on code bit $c_{i,j}$. $\bm{\Psi}_{i,:}^{\mathsf{c},(\ell-1)}$ is the input to BDD, $\bar \mu^{\row,(\ell)}_{i,j}$ is the output of BDD, $l_{i,j}$ is the channel LLR, $\hat l_{i,j}$ is sign of channel LLR, $\tilde \mu^{\row,(\ell)}_{i,j}$ is the output of LUT given in Table~\ref{Tabcomp}, $\psi^{\row,(\ell)}_{i,j}$ is the output of iBDD-CR  and $\tilde l^{\row,(\ell)}_{i,j}$ is the LLR of $\psi^{\row,(\ell)}_{i,j}$.}  
	\label{SysPCCST} 
\end{figure*} 
\section{\GL{Iterative Bounded Distance Decoding with Combined Reliability}}\label{iBDD-SRPC}  

Recently, a variant of iBDD called iBDD-SR has been proposed, where the BDD  outbound messages are modified based on the channel reliabilities to reduce  miscorrections. A detailed explanation of iBDD-SR can be found in \cite{sheikhTCOM19}. We highlight that iBDD-SR \GL{exploits an estimate of the} reliability of the BDD decisions \GL{to produce}  weighted sums of BDD outputs and channel \GL{LLRs} (see \cite[Fig.~2]{sheikhTCOM19}). \GL{We provide next an improved combining rule, based on an enhanced model of the component code decoder behavior. The new rule will be shown to be superior to the weighted sum approach of \cite{sheikhTCOM19}, in terms of both asymptotic decoding threshold and finite-length performance.}

\AG{Consider the decoding of an $(n^2,k^2)$ PC and} let $\bm{\Psi}^{\mathsf{c},(\ell-1)}=[\psi_{i,j}^{\mathsf{c},(\ell-1)}]$ be the decoding result of the $\nc$ column codes at iteration $\ell-1$, \AG{where} $\psi_{i,j}^{\mathsf{c},(\ell-1)}$ corresponds to the decision on code bit $c_{i,j}$. The input of the row decoders at iteration $\ell$ is $\bm{\Psi}^{\mathsf{c},(\ell-1)}$. Without loss of generality, we assume the decoding of the $i$th row code at iteration $\ell$, hence, the input of the decoder is $\bm{\Psi}^{\mathsf{c},(\ell-1)}_{i,:}$.\footnote{\SH{Recall our notation where $\boldsymbol{X}_{i,:}$ denotes the $i$-th row of $\boldsymbol{X}$.}} 
The block diagram of the iBDD-CR algorithm corresponding to the $i$th row decoding at iteration $\ell$ is schematized in Fig.~\ref{SysPCCST}. Let $\bar\mu_{i,j}^{\mathsf r}$ denote the output of \AG{the $i$th row BD decoder}   corresponding to code bit $c_{i,j}$. $\bar\mu_{i,j}^{\mathsf r}$ takes values on a ternary alphabet, $\bar\mu_{i,j}^{\mathsf r} \in \{\pm1, 0 \}$, where the decoded bits are mapped according to $0\mapsto +1$ and $1\mapsto -1$ if BDD is successful, and the output is $0$ in the case of a decoding failure. Let $\tilde{l}_{i,j}^{\mathsf r, (\ell)}$ be the \GL{soft value of} code bit $c_{i,j}$ at iteration $\ell$, which is formed based on the values of the BDD output and the channel LLRs, i.e., $\bar \mu _{i,j}^{\row,(\ell )}$ and $l_{i,j}$, respectively. In Sec.~\ref{sec:DE_PCs}, we derive a closed-form expression for $\tilde{l}_{i,j}^{\mathsf r, (\ell)}$ \AG{via a DE} analysis.\footnote{\GL{Note that the expression is derived under the assumption of an asymptotically-large block length, under an extrinsic message passing variation of the algorithm, as will be discussed in Sec.~\ref{sec:DE_PCs}}.} For now, assume that \GL{\emph{combining}}  $\bar \mu _{i,j}^{\mathsf r,(\ell )}$ and $l_{i,j}$ results in \GL{a soft value} $\tilde{l}_{i,j}^{\mathsf r, (\ell)}$. 
Then, the hard decision on the code bit $c_{i,j}$ produced by the $i$th row decoder is formed as
\begin{equation}\label{eq:BDDchrel_VN}
\psi_{i,j}^{\mathsf{r},(\ell)}=
\mathsf{B}(\tilde{l}_{i,j}^{\mathsf r, (\ell)}) , 
\end{equation}
where ties can be broken with any policy (see Sec.~\ref{sys_mod} for the definition of $\BB(\cdot)$).

The hard decision $\psi_{i,j}^{\mathsf{r},(\ell)}$ is the message
passed from the $i$-th row \AG{decoder} to the $j$-th column \AG{decoder}. In
particular, after applying this procedure to all row \AG{decoders}, the matrix
$\boldsymbol{\Psi}^{\mathsf{r},(\ell)}=[\psi_{i,j}^{\mathsf{r},(\ell)}]$
is formed and used as the input for the $n$ column decoders, and column decoding based on
$\boldsymbol{\Psi}^{\mathsf{r},(\ell)}$ is performed. Assume the decoding of the $j$th column code at iteration $\ell$, hence, the input of the decoder is $\bm{\Psi}^{\mathsf{r},(\ell)}_{:,j}$. As before, we assume that the output of the \AG{$j$th column BD decoder} corresponding to code bit $c_{i,j}$, denoted by $\bar\mu_{i,j}^{\mathsf c}$, takes values on $\{\pm1, 0 \}$.  Similar to row decoding, $\tilde{l}_{i,j}^{\mathsf c, (\ell)}$ is formed based on combining  $\bar\mu_{i,j}^{\mathsf c}$ and $l_{i,j}$. Then, the hard decision on code bit $c_{i,j}$ produced by the $j$th column decoder is formed as 
\begin{equation}\label{eq:BDDchrel_VN_scale}
	\psi_{i,j}^{\mathsf{c},(\ell)}=\BB(\tilde{l}_{i,j}^{\mathsf c, (\ell)}).
\end{equation}
The matrix $\bm{\Psi}^{\mathsf{c},(\ell)}=[\psi_{i,j}^{\mathsf{c},(\ell)}]$ is passed to the $\nc$ row decoders for the decoding iteration $\ell+1$.
The iterative process continues until a maximum number of iterations is reached. 

\GL{With reference to Fig.~\ref{SysPCCST}, observe that in the proposed algorithm the component code decoders exchange only \SH{(binary)} hard decisions. Hence, the contribution of the messages passed among component  decoders to the overall internal decoder data flow} \GL{is comparable to that of conventional iBDD \cite[Sec.~III.A]{staircase_frank}.} 

\section{\AG{Density Evolution} Analysis of iBDD-CR for GLDPC Code Ensembles}\label{sec:DE_PCs}

\subsection{\AG{Density Evolution} Analysis for the bi-AWGN Channel}\label{sec:DE_BIAWGN}

\begin{figure*}[t] \centering 
	\includegraphics[scale=1.4]{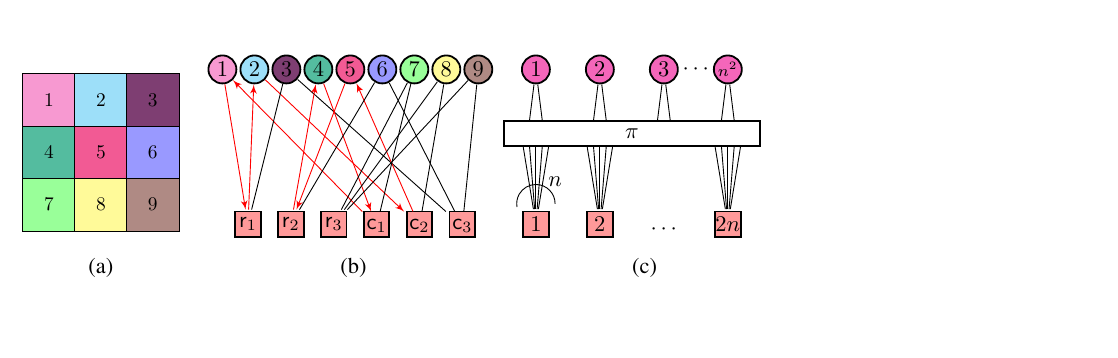}
	\vspace{-1ex}
	\caption{(a) The PC code array with $9$ bits. (b) Tanner graph of the PC. $\mathsf{r_1}$, $\mathsf{r_2}$, and $\mathsf{r_3}$ ($\mathsf{c_1}$, $\mathsf{c_2}$, and $\mathsf{c_3}$) stand for the first, second, and third BCH row (column) \GL{constraint nodes}. \GL{A length-$8$ cycle is highlighted in the Tanner graph. (c) Tanner graph of a generic regular GLDPC code with $n^2$ degree-$2$ VNs and $2n$ degree-$n$ CNs.}}  
	\label{PCgraph} 
\end{figure*}

\GL{We provide in this section a DE analysis of PCs, under iBDD-CR decoding. More specifically, we analyze PCs as members of regular GLDPC code ensembles \cite{Lentmaier98:GLDPC}. To do so, we first recall the  Tanner graph representation of a PC \cite{Tanner1981}. A two-dimensional PC defined by an $(n,k)$ component code (used for both rows and columns of the codeword array) can be represented by a graph consisting of two sets of nodes: $n^2$ degree-$2$ variable nodes (VNs) and $2n$ degree-$n$ constraint nodes (CNs). Each VN is associated to a codeword bit, and each CN is associated to a row/column code. VNs and CNs are then connected by edges according to the constraints defined by the PC construction. An example is provided in Fig.~\ref{PCgraph} for the case \AG{of $n=3$ component code}.}

\GL{PCs can be seen as a special class of GLDPC codes \AG{for which the connections between VNs and CNs are directly defined by the PC structure}. The deterministic nature of the  Tanner graph of PCs has a major consequence in terms of DE analysis. Indeed, DE can applied to graphs whose nodes have a tree-like neighborhood down to a certain depth (which is directly related to the number of \AG{decoding iterations}). To determine the limiting behavior in terms of iterative decoding threshold, the number of iterations (and, consequently, the depth for which a tree-like neighborhood is required) has to be taken to infinity. This demands for the block length to grow large (i.e., in the limit, to infinity). However, even if one would be able to construct a \emph{sequence} of PCs with increasing block length, the requirement of a tree-like neighborhood down to a growing depth cannot be attained by any PC. The girth of the Tanner graph of a PC is, in fact, $8$ (\SH{a length-$8$} cycle in the Tanner graph of the PC of Fig.~\ref{PCgraph}(b) is highlighted), jeopardizing the possibility to perform a proper DE analysis.\footnote{\GL{A notable exception, where a DE analysis of PCs can be performed, was described in \cite{Haeger2017tit} for the binary erasure channel for the limiting case where the code rate tends to $1$}.} Hence, rather than analyzing a specific PC, we resort to the analysis of the GLDPC ensemble encompassing the PC.} 
	
\GL{\AGc{
\SHm{Fig. \ref{PCgraph}(c) depicts the Tanner graph of a generic PC with $n^2$ degree-$2$ VNs and $2n$ degree-$n$ CNs, where the permutation of the edges is represented by the edge interleaver $\pi$. The PC is defined by a particular permutation. Note also that this Tanner graph corresponds to that of a regular GLDPC code for which all CNs are associated the same $(n,k)$ component code.
The set of codes defined by all possible edge permutations yields the GLDPC code ensemble, which contains among its members the two-dimensional \SH{PC} based on the $(n,k)$ component code.}}
To perform a  DE analysis, we  will need to untangle the number of VNs and CNs from the component code block length: The regular GLDPC code ensembles on interest will be hence defined by the CNs component code, the degree of the VNs (which is $2$), and the block length (i.e., the number of VNs).} 



\GL{\begin{remark}
	The Tanner graph of a product/GLDPC code can be used to describe the message-passing schedule between VNs and CNs. Typically, GLDPC ensembles are analyzed under the assumption of a flooding schedule, i.e., at each iteration VNs pass a message to all the neighboring CNs, and all CNs pass a message to their neighboring VNs. While the principle can be applied to PCs, too, in the following we will analyze a message passing schedule that follows from the decoding algorithm described in Sec.~III. To this aim, we divide the CNs into two classes (or types): The class of CNs associated to row decoders and the class of CNs associated to column decoders. The decoding schedule, hence, will involve two half-iterations: In the first half-iteration, only CNs associated to column decoders are active, while in the second half-iteration only CNs associated to row decoders are active.
\end{remark}}

\GL{\begin{remark}
		A further difference in the analysis with respect to the classical analysis of GLDPC code ensembles stems from the DE analysis approach proposed in \cite{JianPfister2017} for GLDPC code ensembles under iBDD. Typically, a VN in a GLDPC code Tanner graph gets as input the corresponding codeword bit channel observation, and it forwards to the neighboring CNs an updated belief. The updated belief accounts for the channel observation and the extrinsic information provided at the previous iteration, by the CNs output. As suggested in \cite{JianPfister2017}, when analyzing the iBDD algorithm it is particularly convenient to provide the channel observations as input to each CN directly, and to let VNs act as simple message routers (i.e., they forward the message received along one edge over the other edge without performing any modification). We adopt this approach in the following. It follows that the combining of the channel soft information with the output of the local BD decoders takes place within the CNs, as emphasized in Fig.~\ref{SysPCCST}.
\end{remark}}

\medskip

\GL{In order to proceed with the DE analysis, a further important aspect needs to be addressed. In particular, the decoding rule specified for the component code decoders in Sec.~III has to be modified in order to render the message passing \emph{extrinsic}.}
From \eqref{eq:BDDchrel_VN}--\eqref{eq:BDDchrel_VN_scale}, one can infer that the standard row (column) decoding of PCs using iBDD-CR does not fall into the extrinsic decision rule category, as the input of row (column) decoder for iteration $\ell$ is the output of column (row) decoder from the iteration $\ell-1$. Therefore, \AG{for analysis purposes, similar to \cite{JianPfister2017,sheikhTCOM19} we modify the algorithm} such that the BDD of the component code is substituted by \GL{an extrinsic rule relying on BDD} \cite[Sec.~II.B]{JianPfister2017}. 
 
 \GL{T}he iBDD-CR decoding \GL{algorithm} with extrinsic BDD of the component codes is explained in the following. \SH{Without loss of generality}, we consider the decoding of the $i$-th row of the PC \SH{at iteration $\ell$}, corresponding to the transmitted component code codeword $\lc=(c_1,\ldots,c_{\nc})$, where the input for the $i$-th row decoder is $\bm{\Psi}_{i,:}^{\mathsf{c},(\ell-1)}=(\psi_{i,1}^{\mathsf{c},(\ell-1)},\ldots,\psi_{i,n}^{\mathsf{c},(\ell-1)})$. For the decision on the code bit $c_{i,j}$, the $j$-th component of $\bm{\Psi}_{i,:}^{\mathsf{c},(\ell-1)}$ is substituted by the channel output $r_{i,j}$, and hence, \SH{$\tilde{\boldsymbol{\psi}}  \buildrel \Delta \over = 
(\psi_{i,1}^{\mathsf{c},(\ell-1)},\ldots,r_{i,j},\ldots,\psi_{i,n}^{\mathsf{c},(\ell-1)})$ is used as the input for BDD. After performing extrinsic BDD, the outbound message on code bit $c_{i,j}$ ($\bar\mu_{i,j}^{\mathsf r}$)  is}
\SH{
\begin{equation}
\label{eq:BDD_VN_extrinsic}
\scalemath{0.912}{	\bar\mu_{i,j}^{\mathsf r} = \begin{cases}
	{(-1)}^{c_i} & \text{if } \ham(\lc, \tilde{\boldsymbol{\psi}})  \leq t \\
	{(-1)}^{\tilde{c}_i} & \text{if } \ham(\lc, \tilde{\boldsymbol{\psi}}) > t \text{ and
	} \exists{\tilde{\lc}} \; \text{such that} \;\ham(\tilde{\lc}, \tilde{\boldsymbol{\psi}}) \leq t \\
	0 & \text{otherwise}
	\end{cases},}
\end{equation}
where $\tilde{\lc}=(\tilde{c}_1,\ldots,\tilde{c}_{\nc})$ is a valid component codeword. The same decoding rule can be employed for iBDD-CR decoding with extrinsic BDD of column codes.}

\begin{remark}\label{rr0}
The extrinsic decoding rule \AG{requires decoding $n$ times each  component code}, which is  complex for a practical system. We highlight that we only use the extrinsic decoding rule to derive the DE \GL{analysis}, whereas for the performance evaluation via simulations we use the more practical decoder outlined in Sec.~\ref{sys_mod}, i.e., the standard (intrinsic) row/column decoding of the component codes is employed.  
\end{remark}


We consider transmission over the bi-AWGN channel where a length-$n$ BCH component code with error-correcting capability $t$ is assumed at the CNs. 
We denote by $p_\mathsf{ch}$ the channel output error probability yielded by applying hard detection to the bi-AWGN channel output\GL{, i.e.,} $p_\mathsf{ch}=\Q\left(\frac{1}{\sigma}\right)$. We consider two sets of equal size for CNs, where each set defines a CN type. We refer to the two CN types as row and column CN types. Each VN is connected to one row-type CN and to one column-type CN. Each decoding iteration consists of one row CN \GL{processing}, followed by one column CN \GL{processing}. In the following, we denote by $\mep$ the error probability associated to the messages exchanged by the component decoders \GL{(via VNs)}. In particular, we denote by $\mep^{\row,(\ell)}$ and $\mep^{\col,(\ell)}$ the message error probability at the output of the row component decoder (row-type CN) and column component decoder (column-type CN), respectively, at the $\ell$th iteration. The message error probability at the input of a row-type CN at the $\ell$th iteration is given by $\mep^{\col,(\ell-1)}$, whereas the message error probability at the input of a column-type CN during the $\ell$th iteration is $\mep^{\row,(\ell)}$.

Without loss of generality,  \AG{consider the row-type CN operation at iteration $\ell$ of \GL{DE}}.
The combining yields \GL{a soft estimate for} code bit $c_{i,j}$, i.e., ${\opt}_{i,j}^{\mathsf r,(\ell)}$, given the corresponding BDD output and channel LLR are $\bar{\mu}_{i,j}^{\mathsf r,(\ell)}$ and $l_{i,j}$, respectively. By applying Bayes' rule to the definition of ${\opt}_{i,j}^{\mathsf r,(\ell)}$, \GL{we have}
\GL{\begin{align}
\opt_{i,j}^{\row,(\ell )} & \buildrel \Delta \over = \ln \frac{{p\left( {\bar \mu _{i,j}^{\row,(\ell )},{{l}_{i,j}}|{c_{i,j}} = 0} \right)}}{{p\left( {\bar \mu _{i,j}^{\row,(\ell )},{{l}_{i,j}}|{c_{i,j}} = 1} \right)}} \nonumber \\ & = \ln \frac{{p\left( {\bar \mu _{i,j}^{\row,(\ell )}|{{l}_{i,j}},{c_{i,j}} = 0} \right)}}{{p\left( {\bar \mu _{i,j}^{\row,(\ell )}|{{l}_{i,j}},{c_{i,j}} = 1} \right)}} + \ln \frac{{p\left( {{{l}_{i,j}}|{c_{i,j}} = 0} \right)}}{{p\left( {{{l}_{i,j}}|{c_{i,j}} = 1} \right)}} \nonumber\\
&=\ln \frac{{p\left( {\bar \mu _{i,j}^{\row,(\ell )}|{l_{i,j}},{c_{i,j}} = 0} \right)}}{{p\left( {\bar \mu _{i,j}^{\row,(\ell )}|{l_{i,j}},{c_{i,j}} = 1} \right)}} + l_{i,j}. \label{LLRoutnosimp} 
\end{align} 
where $l_{i,j}=\frac{2}{\sigma^2}y_{i,j}$. \AG{It follows  $l_{i,j}\sim\mathcal{N}(2/\sigma^2,4/\sigma^2)$ if $c_{i,j}=0$ and $l_{i,j}\sim\mathcal{N}(-2/\sigma^2,4/\sigma^2)$ if $c_{i,j}=1$}.} 

In general, computing the first term of \eqref{LLRoutnosimp} is a formidable task, since the extrinsic decoding rule \eqref{eq:BDD_VN_extrinsic} results in a statistical dependence between $\bar \mu _{i,j}^{\row,(\ell )}$ and $l_{i,j}$. \AG{The trick for computing  this term is that} $\bar \mu _{i,j}^{\row,(\ell )}$ only depends on the \GL{hard value} $\hat l_{i,j}$ and not \GL{on the reliability} $|l_{i,j}|$.\footnote{\GL{Recall that, for what concerns the computation of $\opt_{i,j}^{\row,(\ell )}$} in the extrinsic message passing decoding rule, the input of the row decoder corresponding to code bit $c_{i,j}$ is \GL{given} by $r_{i,j}=\BB(l_{i,j})$. As $\BB(\cdot)$ is a mapper operating on the sign of its input, $\bar \mu _{i,j}^{\row,(\ell )}$ only depends on $\hat l_{i,j}$.} In fact, the extrinsic decoding rule imposes that $l_{i,j} \to \hat l_{i,j} \to \bar \mu _{i,j}^{\row,(\ell )}$ forms a Markov chain, i.e., given $\hat l_{i,j}$, \SH{$\bar \mu _{i,j}^{\row,(\ell )}$} and $l_{i,j}$ are statistically independent. Therefore, one can \GL{re-state} \eqref{LLRoutnosimp} as
\begin{align}\label{LLRoutv1} 
\opt_{i,j}^{\row,(\ell )} = \mut_{i,j}^{\row,(\ell )} + l_{i,j}
\end{align}      
where $\mut_{i,j}^{\row,(\ell )}$ is defined as
\begin{align}\label{mutild_r} 
\mut_{i,j}^{\row,(\ell )}\triangleq\ln \frac{{p\left( {\bar \mu _{i,j}^{\row,(\ell )}|{\hat l_{i,j}},{c_{i,j}} = 0} \right)}}{{p\left( {\bar \mu _{i,j}^{\row,(\ell )}|{\hat l_{i,j}},{c_{i,j}} = 1} \right)}}.
\end{align}  
Similarly, for the column-type CN operation the \GL{soft value} of the VN corresponding to code bit $c_{i,j}$ is derived as 
\begin{align}\label{LLRoutv1_col} 
\opt_{i,j}^{\col,(\ell )} = \mut_{i,j}^{\col,(\ell )} + l_{i,j}, 
\end{align}    
where $\mut_{i,j}^{\col,(\ell )}$ is defined as
\begin{align}\label{mutild_c} 
\mut_{i,j}^{\col,(\ell )}\triangleq\ln \frac{{p\left( {\bar \mu _{i,j}^{\col,(\ell )}|{\hat l_{i,j}},{c_{i,j}} = 0} \right)}}{{p\left( {\bar \mu _{i,j}^{\col,(\ell )}|{\hat l_{i,j}},{c_{i,j}} = 1} \right)}}.
\end{align} 

\begin{figure*}[t]
	\vspace*{-20pt}
	\hrulefill		
	\begin{align}\label{xrl}
	& \nonumber \mep^{\row,(\ell)} = \scalemath{0.8}{f^{\Pue}(\mep^{\col,(\ell-1)}){\Q\left( { \frac{\sigma}{2}\text{min}\left(\ln \left( \frac{f^{\Pue}(\mep^{\col,(\ell-1)})}{f^{\Quc}(\mep^{\col,(\ell-1)})}\right),0\right)+\frac{1}{\sigma}} \right) } + f^{\Puc}(\mep^{\col,(\ell-1)}){\Q\left( { \frac{\sigma}{2}\text{min}\left(\ln \left( \frac{f^{\Puc}(\mep^{\col,(\ell-1)})}{f^{\Que}(\mep^{\col,(\ell-1)})}\right),0\right)+\frac{1}{\sigma}} \right) }} + \\ \nonumber &  \scalemath{0.85}{f^{\Puep}(\mep^{\col,(\ell-1)}){\Q\left( { \frac{\sigma}{2}\text{min}\left(\ln \left(\frac{f^{\Puep}(\mep^{\col,(\ell-1)})}{f^{\Quep}(\mep^{\col,(\ell-1)})}\right),0\right)+\frac{1}{\sigma}} \right) } + f^{\Que}(\mep^{\col,(\ell-1)})\left( {\Q\left( { \frac{\sigma}{2}\ln \left(\frac{f^{\Que}(\mep^{\col,(\ell-1)})}{f^{\Puc}(\mep^{\col,(\ell-1)})}\right)+\frac{1}{\sigma}} \right)}-\Q\left(\frac{1}{\sigma}\right) \right) \cdot \bar{\mathbbm{U}} \left( \ln \left(\frac{f^{\Que}(\mep^{\col,(\ell-1)})}{f^{\Puc}(\mep^{\col,(\ell-1)})}\right) \right)} \\ \nonumber & + \scalemath{0.85}{f^{\Quc}(\mep^{\col,(\ell-1)})\left( {\Q\left( { \frac{\sigma}{2}\ln \left(\frac{f^{\Quc}(\mep^{\col,(\ell-1)})}{f^{\Pue}(\mep^{\col,(\ell-1)})}\right)+\frac{1}{\sigma}} \right)}-\Q\left(\frac{1}{\sigma}\right) \right) \cdot \bar{\mathbbm{U}} \left( \ln \left(\frac{f^{\Quc}(\mep^{\col,(\ell-1)})}{f^{\Pue}(\mep^{\col,(\ell-1)})}\right) \right)} \\  & + \scalemath{0.85}{f^{\Quep}(\mep^{\col,(\ell-1)})\left( {\Q\left( { \frac{\sigma}{2}\ln \left(\frac{f^{\Quep}(\mep^{\col,(\ell-1)})}{f^{\Puep}(\mep^{\col,(\ell-1)})}\right)+\frac{1}{\sigma}} \right)}-\Q\left(\frac{1}{\sigma}\right) \right) \cdot \bar{\mathbbm{U}} \left( \ln \left(\frac{f^{\Quep}(\mep^{\col,(\ell-1)})}{f^{\Puep}(\mep^{\col,(\ell-1)})}\right) \right)}
	\end{align}
	\hrulefill
\end{figure*}

\begin{figure*}[t]
	\vspace*{-20pt}
	\hrulefill		
	\begin{align}\label{xcl}
	& \nonumber \mep^{\col,(\ell)}  = \scalemath{0.85}{f^{\Pue}(\mep^{\row,(\ell)}){\Q\left( { \frac{\sigma}{2}\text{min}\left(\ln \left( \frac{f^{\Pue}(\mep^{\row,(\ell)})}{f^{\Quc}(\mep^{\row,(\ell)})}\right),0\right)+\frac{1}{\sigma}} \right) } + f^{\Puc}(\mep^{\row,(\ell)}){\Q\left( { \frac{\sigma}{2}\text{min}\left(\ln \left( \frac{f^{\Puc}(\mep^{\row,(\ell)})}{f^{\Que}(\mep^{\row,(\ell)})}\right),0\right)+\frac{1}{\sigma}} \right) }} + \\ \nonumber &  \scalemath{0.85}{f^{\Puep}(\mep^{\row,(\ell)}){\Q\left( { \frac{\sigma}{2}\text{min}\left(\ln \left(\frac{f^{\Puep}(\mep^{\row,(\ell)})}{f^{\Quep}(\mep^{\row,(\ell)})}\right),0\right)+\frac{1}{\sigma}} \right) } + f^{\Que}(\mep^{\row,(\ell)})\left( {\Q\left( { \frac{\sigma}{2}\ln \left(\frac{f^{\Que}(\mep^{\row,(\ell)})}{f^{\Puc}(\mep^{\row,(\ell)})}\right)+\frac{1}{\sigma}} \right)}-\Q\left(\frac{1}{\sigma}\right) \right) \cdot \bar{\mathbbm{U}} \left( \ln \left(\frac{f^{\Que}(\mep^{\row,(\ell)})}{f^{\Puc}(\mep^{\row,(\ell)})}\right) \right)} \\ \nonumber & + \scalemath{0.85}{f^{\Quc}(\mep^{\row,(\ell)})\left( {\Q\left( { \frac{\sigma}{2}\ln \left(\frac{f^{\Quc}(\mep^{\row,(\ell)})}{f^{\Pue}(\mep^{\row,(\ell)})}\right)+\frac{1}{\sigma}} \right)}-\Q\left(\frac{1}{\sigma}\right) \right) \cdot \bar{\mathbbm{U}} \left( \ln \left(\frac{f^{\Quc}(\mep^{\row,(\ell)})}{f^{\Pue}(\mep^{\row,(\ell)})}\right) \right)} \\  & + \scalemath{0.85}{f^{\Quep}(\mep^{\row,(\ell)})\left( {\Q\left( { \frac{\sigma}{2}\ln \left(\frac{f^{\Quep}(\mep^{\row,(\ell)})}{f^{\Puep}(\mep^{\row,(\ell)})}\right)+\frac{1}{\sigma}} \right)}-\Q\left(\frac{1}{\sigma}\right) \right) \cdot \bar{\mathbbm{U}} \left( \ln \left(\frac{f^{\Quep}(\mep^{\row,(\ell)})}{f^{\Puep}(\mep^{\row,(\ell)})}\right) \right)}
	\end{align}
	\hrulefill
\end{figure*} 
\GL{The derivation of $\mut_{i,j}^{\row,(\ell )}$ and $\mut_{i,j}^{\col,(\ell )}$ under extrinsic iBDD-CR decoding is provided by the next proposition.}

\medskip


\begin{proposition}\label{p1lab}
\GL{Over the bi-AWGN channel, \SH{the values of}} $\mut_{i,j}^{\row,(\ell )}$ and $\mut_{i,j}^{\col,(\ell )}$ are provided in Table~\ref{Tabcomp} and Table~\ref{Tabcompc}, respectively,  \GL{where the} message error probability at the row-type and column-type CN at the $\ell$th iteration is given by \eqref{xrl} and \eqref{xcl}, respectively, with $\mep^{\col,(0)}=p_\mathsf{ch}$, \GL{and the} values of $f^{\Pue}(\mep)$, $f^{\Puc}(\mep)$, $f^{\Que}(\mep)$, $f^{\Quc}(\mep)$, $f^{\Puep}(\mep)$, $f^{\Quep}(\mep)$ in Tables~\ref{Tabcomp}--\ref{Tabcompc} are derived in \eqref{pe}--\eqref{q_ep}.
\end{proposition}

\GL{The proof of Proposition \ref{p1lab} is given in Appendix~\ref{APP1}. Owing to \eqref{LLRoutnosimp}, we remark that the combining rule is optimal for GLDPC code ensembles in the sense of minimizing the message error probability, under extrinsic message passing decoding, in the limit of infinitely large blocks.}

\begin{table}[t]	
	\caption{$\mut _{i,j}^{\row,(\ell )}$ for row-type CN operation (row decoding for PC) over iteration $\ell$, based on the corresponding BDD output $\mu _{i,j}^{\row,(\ell )}$ and channel \GL{output hard decision} $\hat l_{i,j}$.}
	\centering
	\scalebox{1.2}{	
		\begin{tabular}{ccc}
			\toprule
			$\bar \mu _{i,j}^{\row,(\ell )}$ &
			$\hat l_{i,j}$ &
			$\mut_{i,j}^{\row,(\ell )}$ \\[0.1cm]
			\midrule
			$-1$ & $-1$ & $\ln \frac{f^{\Pue}(\mep^{\col,(\ell-1)})}{f^{\Quc}(\mep^{\col,(\ell-1)})}$ \\[0.3cm]
			$1$  & $-1$ & $\ln \frac{f^{\Puc}(\mep^{\col,(\ell-1)})}{f^{\Que}(\mep^{\col,(\ell-1)})}$ \\ [0.3cm]
			$0$ & $-1$ & $\ln \frac{f^{\Puep}(\mep^{\col,(\ell-1)})}{f^{\Quep}(\mep^{\col,(\ell-1)})}$  \\ [0.3cm]
			$-1$  & $1$ & $\ln \frac{f^{\Que}(\mep^{\col,(\ell-1)})}{f^{\Puc}(\mep^{\col,(\ell-1)})}$ \\[0.3cm]
			$1$ & $1$ & $\ln \frac{f^{\Quc}(\mep^{\col,(\ell-1)})}{f^{\Pue}(\mep^{\col,(\ell-1)})}$ \\ [0.3cm]
			$0$  & $1$ & $\ln \frac{f^{\Quep}(\mep^{\col,(\ell-1)})}{f^{\Puep}(\mep^{\col,(\ell-1)})}$ \\[0.3cm]			
			\bottomrule
		\end{tabular}
	}
	\label{Tabcomp}
\end{table}

\begin{table}[t]	
	\caption{$\mut _{i,j}^{\col,(\ell )}$ for column-type CN operation (column decoding for PC) over iteration $\ell$, based on the corresponding BDD output $\mu _{i,j}^{\col,(\ell )}$ and channel \GL{output hard decision} $\hat l_{i,j}$. 
	}
	\centering
	\scalebox{1.2}{	
		\begin{tabular}{ccc}
			\toprule
			$\bar \mu _{i,j}^{\col,(\ell )}$ &
			$\hat l_{i,j}$ &
			$\mut_{i,j}^{\col,(\ell )}$ \\[0.1cm]
			\midrule
			$-1$ & $-1$ & $\ln \frac{f^{\Pue}(\mep^{\row,(\ell)})}{f^{\Quc}(\mep^{\row,(\ell)})}$ \\[0.3cm]
			$1$  & $-1$ & $\ln \frac{f^{\Puc}(\mep^{\row,(\ell)})}{f^{\Que}(\mep^{\row,(\ell)})}$ \\ [0.3cm]
			$0$ & $-1$ & $\ln \frac{f^{\Puep}(\mep^{\row,(\ell)})}{f^{\Quep}(\mep^{\row,(\ell)})}$  \\ [0.3cm]
			$-1$  & $1$ & $\ln \frac{f^{\Que}(\mep^{\row,(\ell)})}{f^{\Puc}(\mep^{\row,(\ell)})}$ \\[0.3cm]
			$1$ & $1$ & $\ln \frac{f^{\Quc}(\mep^{\row,(\ell)})}{f^{\Pue}(\mep^{\row,(\ell)})}$ \\ [0.3cm]
			$0$  & $1$ & $\ln \frac{f^{\Quep}(\mep^{\row,(\ell)})}{f^{\Puep}(\mep^{\row,(\ell)})}$ \\[0.3cm]			
			\bottomrule
		\end{tabular}
	}
	\label{Tabcompc}
\end{table} 
\begin{figure}[!t] \centering 
	\includegraphics[width=\columnwidth]{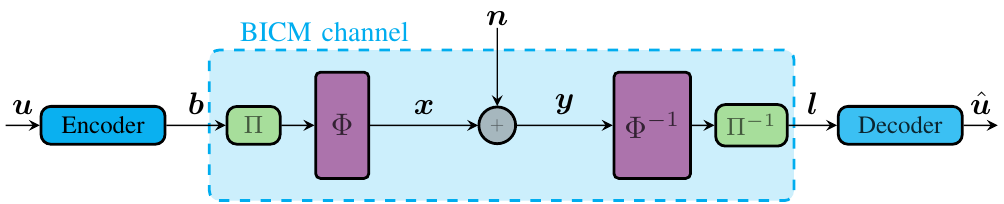}  
	\caption{\GL{Summary of the notation for the BICM scheme}. \SH{The message $\boldsymbol{u}$ is encoded to $\boldsymbol{b}$. After interleaving and mapping, $\boldsymbol{x}$ is sent through a \GL{Gaussian} channel \GL{where the} noise $\boldsymbol{n}$ \GL{is added}, resulting in $\boldsymbol{y}$. \GL{The LLRs corresponding to $\boldsymbol{b}$ are denoted as} $\boldsymbol{l}$, \GL{whereas the decoder output is} $\hat{\boldsymbol{u}}$. $\Phi$ maps $m$ bits to a \GL{modulation} symbol and $\Phi^{-1}$ maps a symbol to $m$ bits.}} 
	\label{BICM} 
	\vspace{-2ex}
\end{figure}

%
%

\begin{figure*}[t]
	\vspace*{-20pt}
	\hrulefill		
	\begin{align}\label{QAMxrl}
	& \scalemath{0.78}{\nonumber \mep^{\row,(\ell)}}  \scalemath{0.8}{ \buildrel \Delta \over = \mathsf{g}(\mep^{\col,(\ell-1)}, \sigma, M) = f^{\Pue}(\mep^{\col,(\ell-1)}){\sum\limits_{j = 0}^{M/2 - 1} {w_j} {\Q\biggl( \frac{{ \text{min}\left(\ln \left( \frac{f^{\Pue}(\mep^{\col,(\ell-1)})}{f^{\Quc}(\mep^{\col,(\ell-1)})}\right),0\right)+\mu_{j}}}{\sigma_{j}} \biggr) } } + f^{\Puc}(\mep^{\col,(\ell-1)}){\sum\limits_{j = 0}^{M/2 - 1} {w_j} {\Q\biggl( \frac{{ \text{min}\left(\ln \left( \frac{f^{\Puc}(\mep^{\col,(\ell-1)})}{f^{\Que}(\mep^{\col,(\ell-1)})}\right),0\right)+\mu_{j}}}{\sigma_{j}} \biggr) } }} + \\ \nonumber &  \scalemath{0.78}{f^{\Puep}(\mep^{\col,(\ell-1)}){\sum\limits_{j = 0}^{M/2 - 1} {w_j} {\Q\biggl( \frac{{ \text{min}\left(\ln \left( \frac{f^{\Puep}(\mep^{\col,(\ell-1)})}{f^{\Quep}(\mep^{\col,(\ell-1)})}\right),0\right)+\mu_{j}}}{\sigma_{j}} \biggr) } } + f^{\Que}(\mep^{\col,(\ell-1)})\biggl( {\sum\limits_{j = 0}^{M/2 - 1} {w_j} \biggl( {\Q\biggl( \frac{{ \ln \left( \frac{f^{\Que}(\mep^{\col,(\ell-1)})}{f^{\Puc}(\mep^{\col,(\ell-1)})}\right)+\mu_{j}}}{\sigma_{j}} \biggr) } - \Q\biggl(\frac{\mu_{j}}{\sigma_{j}} \biggr) \biggr) } \biggr) \cdot \bar{\mathbbm{U}} \biggl( \ln \left(\frac{f^{\Que}(\mep^{\col,(\ell-1)})}{f^{\Puc}(\mep^{\col,(\ell-1)})}\right) \biggr)} \\ \nonumber & + \scalemath{0.78}{f^{\Quc}(\mep^{\col,(\ell-1)})\biggl( {\sum\limits_{j = 0}^{M/2 - 1} {w_j}  \biggl( {\Q\biggl( \frac{{ \ln \left( \frac{f^{\Quc}(\mep^{\col,(\ell-1)})}{f^{\Pue}(\mep^{\col,(\ell-1)})}\right)+\mu_{j}}}{\sigma_{j}} \biggr) } - \Q\left(\frac{\mu_{j}}{\sigma_{j}} \right) \biggr) } \biggr) \cdot \bar{\mathbbm{U}} \left( \ln \left(\frac{f^{\Quc}(\mep^{\col,(\ell-1)})}{f^{\Pue}(\mep^{\col,(\ell-1)})}\right) \right)} \\  & + \scalemath{0.78}{f^{\Quep}(\mep^{\col,(\ell-1)})\biggl( {\sum\limits_{j = 0}^{M/2 - 1} {w_j}  \biggl({\Q\biggl( \frac{{ \ln \biggl( \frac{f^{\Quep}(\mep^{\col,(\ell-1)})}{f^{\Puep}(\mep^{\col,(\ell-1)})}\biggr)+\mu_{j}}}{\sigma_{j}} \biggr) } - \Q\biggl(\frac{\mu_{j}}{\sigma_{j}}\biggr) \biggr) } \biggr( \cdot \bar{\mathbbm{U}} \biggl( \ln \left(\frac{f^{\Quep}(\mep^{\col,(\ell-1)})}{f^{\Puep}(\mep^{\col,(\ell-1)})}\right) \biggr)}
	\end{align}
	\hrulefill
\end{figure*} 

\begin{figure*}[t]
	\vspace*{-20pt}
	\hrulefill		
	\begin{align}\label{QAMxcl}
	& \nonumber \mep^{\col,(\ell)} \buildrel \Delta \over = \mathsf{g}(\mep^{\row,(\ell)}, \sigma, M) = \scalemath{0.8}{f^{\Pue}(\mep^{\row,(\ell)}){\sum\limits_{j = 0}^{M/2 - 1} {w_j} {\Q\biggl( \frac{{ \text{min}\left(\ln \left( \frac{f^{\Pue}(\mep^{\row,(\ell)})}{f^{\Quc}(\mep^{\row,(\ell)})}\right),0\right)+\mu_{j}}}{\sigma_{j}} \biggr) } } + f^{\Puc}(\mep^{\row,(\ell)}){\sum\limits_{j = 0}^{M/2 - 1} {w_j} {\Q\biggl( \frac{{ \text{min}\left(\ln \left( \frac{f^{\Puc}(\mep^{\row,(\ell)})}{f^{\Que}(\mep^{\row,(\ell)})}\right),0\right)+\mu_{j}}}{\sigma_{j}} \biggr) } }} + \\ \nonumber &  \scalemath{0.8}{f^{\Puep}(\mep^{\row,(\ell)}){\sum\limits_{j = 0}^{M/2 - 1} {w_j} {\Q\biggl( \frac{{ \text{min}\left(\ln \left( \frac{f^{\Puep}(\mep^{\row,(\ell)})}{f^{\Quep}(\mep^{\row,(\ell)})}\right),0\right)+\mu_{j}}}{\sigma_{j}} \biggr) } } + f^{\Que}(\mep^{\row,(\ell)})\biggl( {\sum\limits_{j = 0}^{M/2 - 1} {w_j} \biggl( {\Q\biggl( \frac{{ \ln \left( \frac{f^{\Que}(\mep^{\row,(\ell)})}{f^{\Puc}(\mep^{\row,(\ell)})}\right)+\mu_{j}}}{\sigma_{j}} \biggr) } - \Q\left(\frac{\mu_{j}}{\sigma_{j}} \right) \biggr) } \biggr) \cdot \bar{\mathbbm{U}} \left( \ln \left(\frac{f^{\Que}(\mep^{\row,(\ell)})}{f^{\Puc}(\mep^{\row,(\ell)})}\right) \right)} \\ \nonumber & + \scalemath{0.8}{f^{\Quc}(\mep^{\row,(\ell)})\biggl( {\sum\limits_{j = 0}^{M/2 - 1} {w_j} \biggl( {\Q\biggl( \frac{{ \ln \left( \frac{f^{\Quc}(\mep^{\row,(\ell)})}{f^{\Pue}(\mep^{\row,(\ell)})}\right)+\mu_{j}}}{\sigma_{j}} \biggr) } - \Q\left(\frac{\mu_{j}}{\sigma_{j}} \right) \biggr) } \biggr) \cdot \bar{\mathbbm{U}} \left( \ln \left(\frac{f^{\Quc}(\mep^{\row,(\ell)})}{f^{\Pue}(\mep^{\row,(\ell)})}\right) \right)} \\  & + \scalemath{0.8}{f^{\Quep}(\mep^{\row,(\ell)})\biggl( {\sum\limits_{j = 0}^{M/2 - 1} {w_j} \biggl( {\Q\biggl( \frac{{ \ln \left( \frac{f^{\Quep}(\mep^{\row,(\ell)})}{f^{\Puep}(\mep^{\row,(\ell)})}\right)+\mu_{j}}}{\sigma_{j}} \biggr) } - \Q\left(\frac{\mu_{j}}{\sigma_{j}} \right) \biggr) } \biggr) \cdot \bar{\mathbbm{U}} \left( \ln \left(\frac{f^{\Quep}(\mep^{\row,(\ell)})}{f^{\Puep}(\mep^{\row,(\ell)})}\right) \right)}
	\end{align}
	\hrulefill
\end{figure*}

\subsection{Density Evolution Analysis for BICM with $M^{2}$-QAM}\label{QAM_BICM}

\GL{BICM \cite{Zehavi1992,Caire1998} (see Fig.~\ref{BICM}) has become a \emph{de-facto} standard in optical communications \cite{georg_tcom,Buchali_2016} due to the inherent flexibility and implementation simplicity.} 
One can see the $M^{2}$-QAM modulation as a Cartesian product of two amplitude shift keying (ASK) modulations, i.e, the real and imaginary parts of each $M^{2}$-QAM symbol \GL{are} chosen form an $M$-ASK constellation. \GL{More specifically}, for $M=2^{m}$, the constellation points for both real and imaginary parts of the $M^{2}$-QAM symbol are chosen from $\mathcal{X}\triangleq\{(-2^m+1)\cdot \Delta ,...,-\Delta,\Delta,...,(2^m-1)\cdot \Delta\}$ where $\Delta=\sqrt{\frac{3}{2(M^2-1)}}$ is a scaling factor which normalizes the resulting $M^{2}$-QAM constellation energy to $1$. In the following, we only consider transmission of the real part of $M^{2}$-QAM, as the real and imaginary parts can be treated independently for the AWGN channel.  
\SH{We consider binary reflected Gray coding (BRGC) mapping \cite{BRGC}.}
The output of the AWGN channel at time instant $i$ corresponding to ASK symbol $x_{i}$ is given by 
\begin{align}\label{channel_inst1}
y_{i}=x_{i}+n_{i},
\end{align}  
where $x_{i}\in \mathcal{X}$ and $n_{i}\sim \mathcal{N}(0,\sigma^2)$. 
The LLR of the $k$-th bit level of $y_{i}$ is given as
\begin{align} \label{LLR}
l_i^k = \ln \left(\frac{\sum\limits_{a \in \mathcal{S}_k^0} {{e^{ - \frac{{({y_i} - a{)^2}}}{{2\sigma^2}}}}} }{{\sum\limits_{a \in \mathcal{S}_k^1} {{e^{ - \frac{{({y_i} - a{)^2}}}{{2\sigma^2}}}  }} }}\right),\;k=1,\cdots,m
\end{align}  
where $\mathcal{S}_k^0 \subset \mathcal{X}$ and $\mathcal{S}_k^1 \subset \mathcal{X}$ are sets of size \SH{$2^{m}$} ASK symbols with $0$ and $1$ as the $k$th bit of the corresponding BRGC label, respectively. To alleviate the complexity of the LLR computation the well-known max-log approximation is usually used, yielding \cite[Eq.~6]{Viterbi1998}
\begin{align} \label{LLRMaxlog}
l_i^k & \approx \frac{1}{{2{\sigma ^2}}}\left[ {\mathop {\max }\limits_{a \in {\cal S}_l^0} \left\{ {{{-\left( {{y_i} - a} \right)}^2}} \right\} - \mathop {\max }\limits_{a \in {\cal S}_l^1} \left\{ {{{-\left( {{y_i} - a} \right)}^2}} \right\}} \right] \nonumber \\ &= \frac{1}{{2{\sigma ^2}}}\left[ {\mathop {\min }\limits_{a \in {\cal S}_l^1} \left\{ {{{\left( {{y_i} - a} \right)}^2}} \right\} - \mathop {\min }\limits_{a \in {\cal S}_l^0} \left\{ {{{\left( {{y_i} - a} \right)}^2}} \right\}} \right].
\end{align}         

\begin{align} \label{LLRMaxlog_approx}
p\left( {l|b} \right) = \sum\limits_{j = 0}^{\frac{M}{2} - 1} {{w_j}} {\G}({l;\left( { - 1} \right)^b}{\mu _j},\sigma _j^2),
\end{align} 
where $l$ is the LLR corresponding to transmitted bit $b \in \{0,1\}$, ${w_j} = \frac{{2\left( {{2^{m - \left\lceil {{{\log }_2}\left( {j + 1} \right)} \right\rceil }} - 1} \right)}}{{m \cdot M}}$, $\mu_j=\frac{2\Delta^2(j+1)^2}{\sigma^2}$, and $\sigma^2_j=\frac{4\Delta^2(j+1)^2}{\sigma^2}$, respectively. \GL{To proceed with the DE analysis, we assume that the all-zero codeword is transmitted. This leads to the need to symmetrizing the LLR distribution. To achieve this, we resort to the use of channel adapters \cite{Hou2003}.} 
\SH{Denote by $\bar l$ the LLR of the symmetrized BICM channel. The distribution of $\bar l$ is given as \cite[Eq.~(19)]{Ivanov2016}}
\begin{align} \label{LLRMaxlog_approxllr}
p\left( \bar{l} | b\right) =\frac{p\left( {l|b}\right)+p\left( {-l|1-b}\right)}{2}.
\end{align}  

\medskip

\begin{proposition}
\GL{For a BICM scheme} with BRGC \SH{mapping}, the message error probability at the row-type and column-type CN at the $\ell$th iteration is given by \eqref{QAMxrl} and \eqref{QAMxcl}, respectively, with 
\[
\mep^{\col,(0)}=\sum\limits_{j = 0}^{\frac{M}{2} - 1} {{w_j}}\cdot\Q\left(\frac{\mu_j}{\sigma_j}\right).
\]
\end{proposition}
\GL{The proof is given in Appendix~\ref{AB}.}

\section{Numerical  Results}\label{Simres}

In this section, in order to have a direct comparison with iBDD-SR, we consider PCs with the same component codes as those considered in \cite{sheikhTCOM19}, i.e., PCs with ($255$,$231$,$3$) and ($511$,$484$,$3$) BCH component codes.\footnote{Component codes with long block length and 
 $t=3$ are interesting for fiber-optic communications as their error floor 
 is below $10^{-15}$ and the decoder can be efficiently implemented using LUTs \cite[Appendix~I]{staircase_frank}.} The code rates of the PCs are  $R=0.820$ and $0.897$, respectively. \SH{Also, we  consider $12$ decoding iterations.} 

\SH{\begin{remark}\label{remarkapp}
In the case of  errors with high reliability, i.e., LLRs with high magnitude and  wrong sign, the decoding rule in \eqref{eq:BDDchrel_VN}--\eqref{eq:BDDchrel_VN_scale} will be unable to correct such errors, as it is likely that $\BB(\tilde{l}_{i,j}^{\mathsf r, (\ell)})=\BB({l}_{i,j})$. Therefore, similar to  iBDD-SR \cite[Sec.~VI]{sheikhTCOM19}, one should selectively apply iBDD-CR to to increase the chance of correcting such errors. In particular, unless otherwise specified, we consider iBDD-CR and iBDD-SR for some iterations and then we append a few conventional iBDD iterations. The \GL{additional} iBDD iterations increase the chance of correcting transmission errors with high reliability, as the iBDD decoding rule is independent of channel reliabilities. \GL{Specifically}, we consider a maximum of $10$ iBDD-CR (or iBDD-SR) iterations followed by $2$ conventional iBDD iterations. For the sake of fairness, other decoding algorithms are evaluated with $12$ decoding iterations.
\end{remark}}



\begin{figure}[t] \centering 
	\includegraphics[scale=0.82]{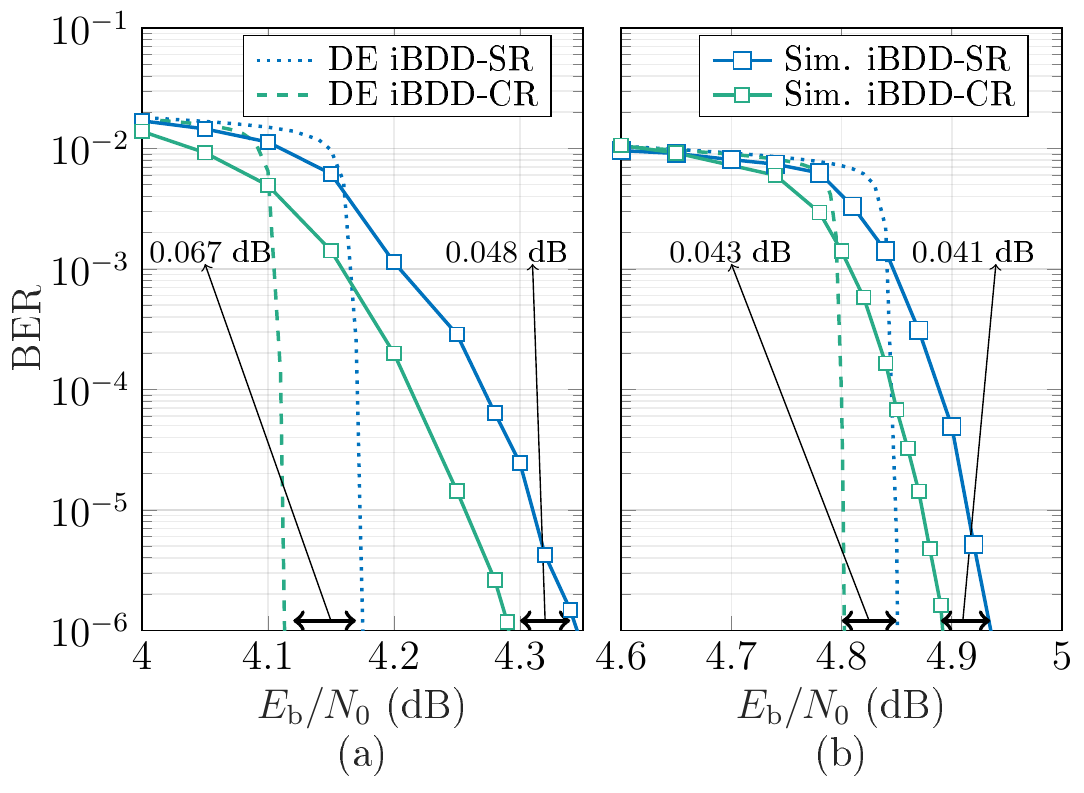}  
	\caption{Comparison between DE \GL{thresholds (computed for the relevant GLDPC ensembles)} and \GL{BER} performance of iBDD-CR and iBDD-SR algorithms \GL{applied to PCs}. The \GL{BCH} component code\GL{s} for (a) and (b) \GL{have parameters} ($255$,$231$,$3$) and ($511$,$484$,$3$), respectively.}  \vspace{-2ex}
	\label{DEsimt3} 
\end{figure}

\begin{figure}[t] \centering 
	\includegraphics[scale=0.6]{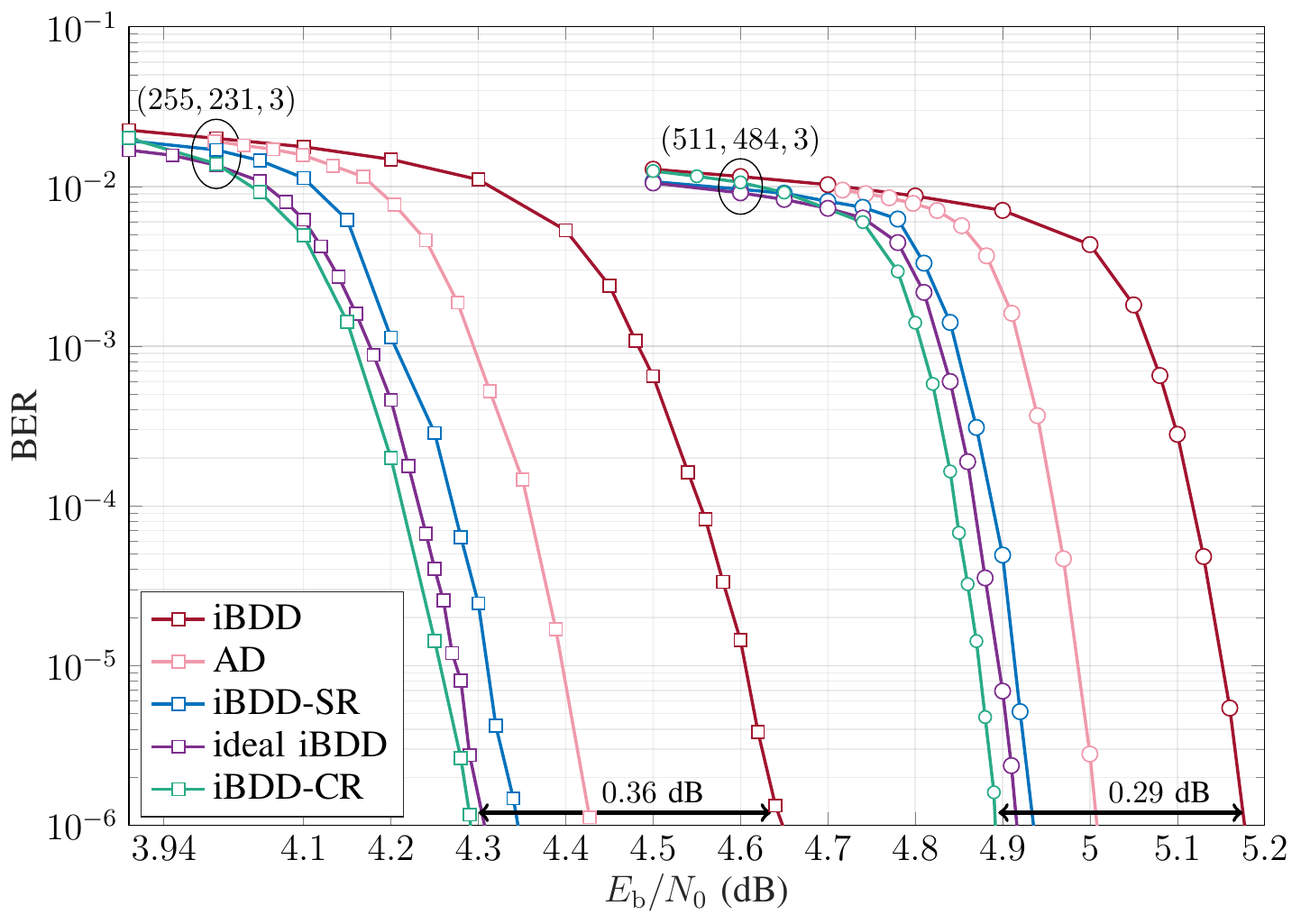}  
	\caption{Performance of iBDD, ideal iBDD, AD, iBDD-SR, and iBDD-CR for a PC with \GL{a ($255$,$231$,$3$) BCH} component code  and a staircase code with \GL{a ($511$,$484$,$3$) BCH} component code.}  \vspace{-2ex}
	\label{percompt3} 
\end{figure} 

\begin{figure*}[t] \centering 
	\includegraphics[scale=0.67]{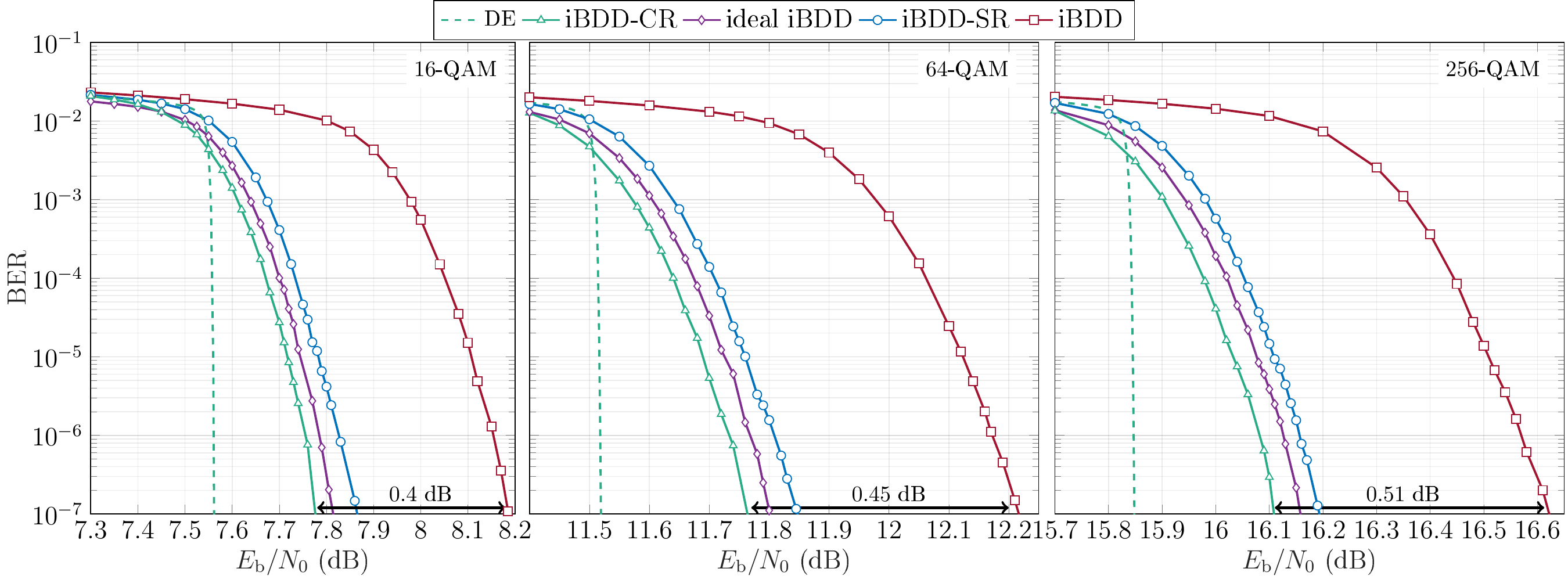}  
	\caption{Performance of iBDD, ideal iBDD, iBDD-SR, and iBDD-CR with \GL{unquantized} LLRs for \GL{a ($255$,$231$,$3$) BCH} component code  in \GL{a BICM scheme} with $16$-QAM, $64$-QAM, and $256$-QAM modulation. 
	}  
	\label{QAM_mod_all_schemes} 
\end{figure*}

\begin{figure}[t] \centering 
	\includegraphics[scale=0.6]{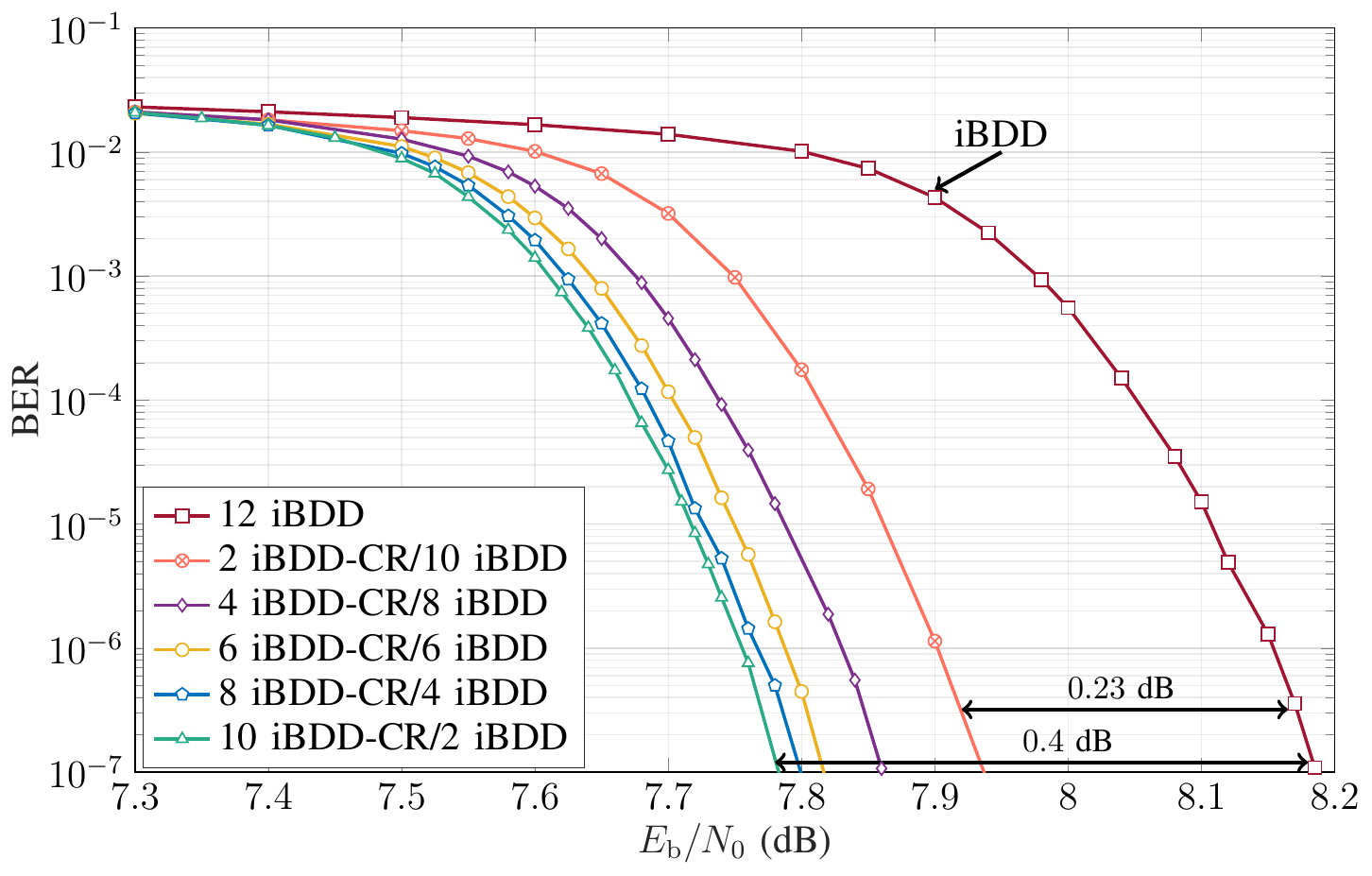}  
	\caption{Performance of iBDD-CR with different appended iBDD iterations for the BICM channel with $16$-QAM modulation and PC with BCH component code ($255$,$231$,$3$). 
	}  \vspace{-2ex}
	\label{tradeoff_RA_BMP} 
\end{figure} 

\SH{We highlight that for the PC performance simulation, we obtain the values of $\mut _{i,j}^{\row,(\ell )}$ and $\mut _{i,j}^{\col,(\ell )}$ from the DE for a single $E_\mathrm{b}/N_0$, corresponding to an SNR point in the waterfall region. We found that changing the operating SNR point in the waterfall region (which in principle results in different values for $\mut _{i,j}^{\row,(\ell )}$ and $\mut _{i,j}^{\col,(\ell )}$) yields a minor performance difference in the simulation results.} Furthermore, in the following simulation results we \GL{restrict to} $\mut _{i,j}^{\row,(\ell )}=\mut _{i,j}^{\col,(\ell )}$, i.e., for a given decoding iteration the values of $\mut _{i,j}^{\row,(\ell )}$ in Table~\ref{Tabcomp} are used for both row and column decoding, as we found that \AG{this induces a negligible performance loss}.

In Fig.~\ref{DEsimt3}, the performance of PCs with iBDD-CR and iBDD-SR over the bi-AWGN channel is shown and compared with the \GL{DE thresholds of the corresponding GLDPC code ensembles}. As \GL{it} can be seen, the performance improvement of iBDD-CR over iBDD-SR is well-predicted by the DE analysis, \GL{confirming that} DE can be used for as a tool for \GL{the} optimization \GL{of the decoding algorithm}. The gap between \GL{the DE thresholds} and \GL{the BER curves} is due to two reasons. \GL{First and foremost,} DE predicts the performance of the GLDPC \GL{code} ensemble with infinite long block length. By increasing the PC component \GL{code} block length from $255$ to $511$, \GL{the} gap is reduced, as \GL{it} can be seen \GL{in} Fig.~\ref{DEsimt3}. \GL{Second,} \GL{DE analysis relies on an} extrinsic message passing \GL{variation of the algorithms}, whereas \GL{the} simulation results \GL{employ} standard (intrinsic) message passing.    

\newcommand{\tablehighlight}{}
\begin{table}[t]	
	\caption{Comparison of iBDD-CR and iBDD for PCs with ($255$,$231$,$3$) and ($511$,$484$,$3$) BCH component codes, with code rates of $0.820$ and $0.897$, respectively. The $E_\mathrm{b}/N_0$ for iBDD and iBDD-CR are measured at $\text{BER} = 10^{-6}$ from the simulations. The corresponding Shannon limits are also shown. 
}
	\centering
	\renewcommand{\arraystretch}{1.2}
	\scalebox{0.96}{	
		\begin{tabular}{ccccc}
			\toprule
			\makecell{component \\ code} &
			\makecell{decoding \\ algorithm} &
			\makecell{\tablehighlight{$E_\mathrm{b}/N_0$} 
				\tablehighlight{[dB]}} &
			\makecell{Shannon limit [dB]}&\\
			\midrule
			($255$,$231$,$3$) & iBDD & $4.62$ & $3.54$ (HD) &   \\
			($255$,$231$,$3$) & iBDD-CR & $4.29$ & $2.23$ (SD)  \\
			($511$,$484$,$3$) & iBDD & $5.18$ &  $4.36$ (HD) &   \\
            ($511$,$484$,$3$) & iBDD-CR & $4.89$ & $3.15$ (SD)  \\			
			\bottomrule
		\end{tabular}
	}
	\label{Tabcomp1}
	\vspace{-4ex}
\end{table}

In Fig.~\ref{percompt3}, we show the performance of iBDD-CR, iBDD, ideal iBDD
, \AG{anchor decoding (AD) \cite{Hag18}}, and iBDD-SR for transmission \SH{over} the bi-AWGN channel. 
One can see that iBDD-CR outperforms all other decoders. In particular, iBDD-CR performs even better than ideal iBDD. The performance gain of iBDD-CR over iBDD is $0.36$ dB and $0.29$ dB for PCs with component codes ($255$,$231$,$3$) and ($511$,$484$,$3$), respectively. \SH{In Table~\ref{Tabcomp1}, we show the required $E_\mathrm{b}/N_0$ for iBDD-CR and iBDD to achieve a BER of $10^{-6}$. We also show the corresponding hard-decision (HD) and soft-decision (SD) Shannon limits. Note that iBDD should be compared with the HD Shannon limit, while iBDD-CR should be compared with the SD counterpart, as the algorithm exploits the channel  LLRs.} The gap between the performance of iBDD-CR and the SD Shannon limit is \AG{mainly due to the fact that the structure of \SH{iBDD-CR} structure is intentionally kept very similar to that of iBDD}, \SH{in order to constrain the decoder complexity and \AG{the internal decoder data flow}.} By allowing the exchange of soft information between component decoders this gap can be closed further at the cost of significantly \AG{higher data flow and  complexity} (see \cite{She18b,She19,YibitTCOM} for more details). 

In Fig.~\ref{QAM_mod_all_schemes}, we show the performance of \AG{a  BICM system  using a PC with component  code ($255$,$231$,$3$) for  iBDD, ideal iBDD, iBDD-SR, and iBDD-CR} and $16$-QAM, $64$-QAM, and $256$-QAM. \SH{We also show the DE threshold for the corresponding GLDPC ensemble.} iBDD-CR outperforms others decoders and the \SH{DE analysis predicts} the performance of iBDD-CR accurately. Furthermore, the gain of iBDD-CR over iBDD improves by increasing the modulation order\AG{; the gain \SH{is} up to} $0.51$ dB for $256$-QAM. 
 

\SH{In Fig.~\ref{tradeoff_RA_BMP}, we examine the effect of appended iBDD iterations on the performance of iBDD-CR. We split the total of $12$ decoding iterations between iBDD-CR and iBDD. As it can be seen, increasing the number of iBDD-CR iterations yields a performance improvement, where such improvement saturates at $10$ iBDD-CR and $2$ iBDD iterations. This motivates the choice of appending $2$ iBDD iterations for performance evaluation of iBDD-CR (see Remark~\ref{remarkapp}).} 
 
\begin{figure}[t] \centering 
	\includegraphics[scale=0.68]{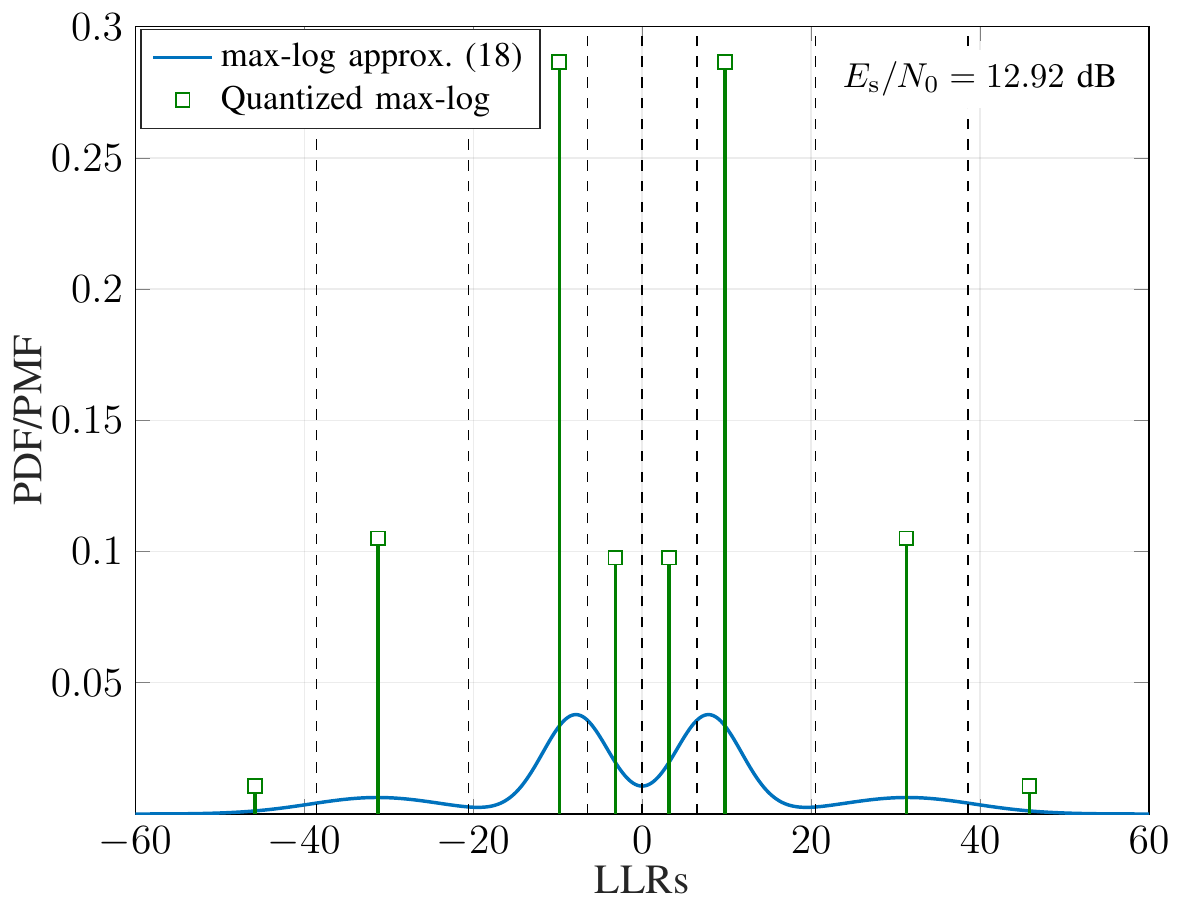}  
	\caption{Comparison between the PDF of the LLRs and the corresponding quantized PMF using $3$ bits for $16$-QAM modulation and $E_\mathrm{s}/N_0=12.92$ dB. The dashed line shows the boundaries of the optimized nonuniform quantization based on the Lloyd-Max algorithm.}  \vspace{-4ex}
	\label{quant_comp} 
\end{figure}

\begin{figure}[t] \centering 
	\includegraphics[scale=0.6]{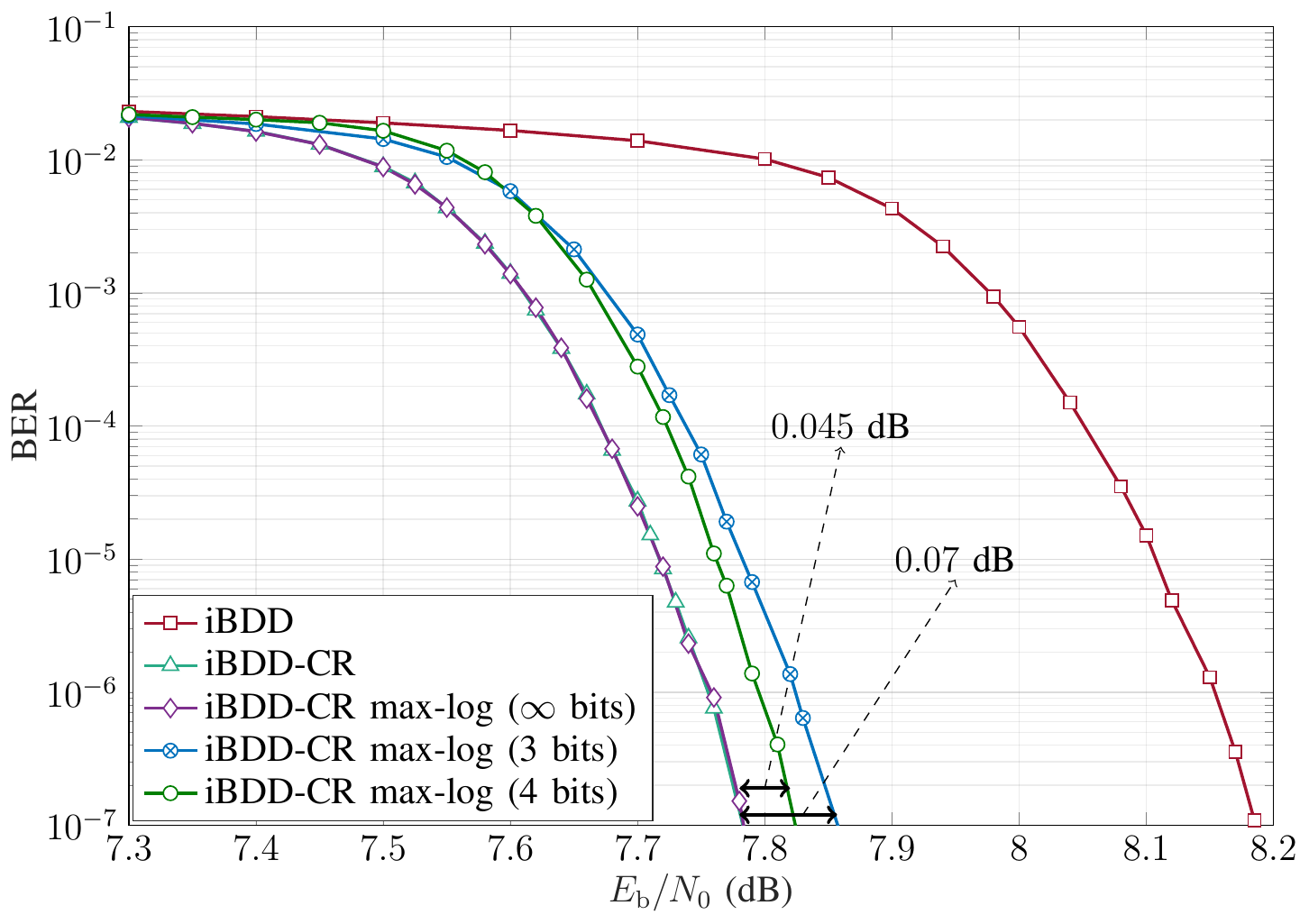}  
	\caption{Performance of iBDD-CR with exact, max-log approximation, and quantized \SH{channel} LLRs for the BICM channel with $16$-QAM modulation and PC with BCH component code ($255$,$231$,$3$).}  
	\label{Max_loyd_alg} 
\end{figure}

\SH{To evaluate the sensitivity of iBDD-CR to channel LLR quantization,
we resort to a classical quantization scheme called \GL{Lloyd–Max} algorithm \cite{Max1960,Lloyd1980}, which \GL{aims at optimizing} the quantization levels in the sense of minimizing the mean squared error between LLRs and the corresponding quantized values.\footnote{We highlight that in this paper our approach is to just show the feasibility of iBDD-CR implementation with limited \SH{channel LLR quantization levels}, using a known quantization technique. In general, one can exploit the properties of the quantized channel in decoding rule to reduce the sensitivity to quantization errors. This analysis is beyond the scope of this paper.}
As an example, Fig.~\ref{quant_comp} shows the distribution of LLRs for $16$-QAM at $E_\mathrm{s}/N_0=12.92$ dB,\footnote{We highlight that $E_\mathrm{s}/N_0=12.92$ dB corresponds to $E_\mathrm{b}/N_0=7.76$ dB which is a point selected in the waterfall region of iBDD-CR (see Fig.~\ref{tradeoff_RA_BMP}).} the optimized quantization values, and the corresponding boundaries using $3$ bits based on the \GL{Lloyd–Max} algorithm. 
As it can be seen, Lloyd–Max yields nonuniform quantization values.} 

\SH{In Fig.~\ref{Max_loyd_alg}, we investigate the effect of \SH{max-log channel LLR approximation and channel LLR quantization based on the Lloyd–Max algorithm on the performance of iBDD-CR}.
In particular, we consider a BICM channel with $16$-QAM modulation, PC with BCH component code ($255$,$231$,$3$), and \AG{$3$-bit and $4$-bit quantization}. \SH{One can see that the performance of iBDD-CR with exact channel LLR computation \eqref{LLR} and max-log channel LLR approximation \eqref{LLRMaxlog} is almost identical}, hence, in a practical system max-log approximation can be employed to reduce the complexity of LLR computations. Furthermore, at a BER of $10^{-7}$, the performance loss of iBDD-CR with $3$ and $4$ bits nonuniform \SH{channel LLR} quantization based on 
the Lloyd-Max algorithm is small, i.e., $0.07$ dB and $0.045$ dB, respectively. This \AG{shows} that iBDD-CR has a low sensitivity to channel LLR quantization.} 


\section{Conclusion}

\AG{We proposed an iterative soft-aided decoding algorithm for PCs, called iBDD-CR}. iBDD-CR exploits the LLRs in the BDD of the component codes and has the same decoder data flow as that of conventional iBDD. \AG{We performed a  DE analysis of  the GLDPC code ensemble containing PCs for both the bi-AWGN and BICM channels under extrinsic message passing}. \AG{From the analysis}, \GL{an accurate estimate of} the reliability of the BDD outbound messages, which is essential for implementing the iBDD-CR was derived. \AG{The proposed algorithm attains gains up to $0.51$ dB over conventional iBDD and outperforms ideal iBDD. We showed} that iBDD-CR has a low sensitivity to quantization errors on the channel LLRs and can be implemented using the low-complexity max-log LLR approximation. Overall, \AG{the low required internal data flow  and  low sensitivity to quantization errors of iBDD-CR} 
makes it an attractive solution for optical communication for $400$G and beyond, where an excellent performance along with stringent constraint on latency and power consumption are required. 

\SHm{
We remark that iBDD-CR requires some extra memory compared to iBDD to store the channel LLRs and the LUTs. The  exact required extra memory  depends on the level of parallelism in implementing row/column decoding and the relative required clock cycles of BDD and combining stages. Therefore, the complexity of iBDD-CR should be investigated via hardware implementation, which is left as future work.}

\section*{Acknowledgment}

The authors would like to thank Dr. Alexios Balatsoukas-Stimming for fruitful discussions about the complexity of iBDD-CR.

\appendices

\section{Proof of \GL{Proposition} 1} \label{APP1}
Let us consider the decoding of row-type CNs. In particular, we first compute $\mut_{i,j}^{\row,(\ell )}$ and then we calculate $\mep^{\row,(\ell)}$. At the first iteration, we have $\mep^{\col,(0)}=p_\mathsf{ch}$, i.e., the input of the row-type CN is initialized with the channel error probability.  
\AG{Let $\hat l_{i,j}$ be}  the RV 
representing the sign of the LLR corresponding to code bit $c_{i,j}$ (see Fig.~\ref{SysPCCST}). To compute $\tilde \mu _{i,j}^{\row,(\ell )}$ (see \eqref{mutild_r}), we should compute the probabilities of $\scalemath{0.9}{p ({\bar \mu _{i,j}^{\row,(\ell )}|{\hat l_{i,j}},{c_{i,j}} = 0} )}$ and $\scalemath{0.9}{p( {\bar \mu _{i,j}^{\row,(\ell )}|{\hat l_{i,j}},{c_{i,j}} = 1})}$.  
Depending on the values of $\hat l_{i,j} \in \{\pm 1\}$ and $\mu _{i,j}^{\row,(\ell )} \in \{0, \pm 1\}$, six different terms for $\scalemath{0.9}{p( {\bar \mu _{i,j}^{\row,(\ell )}|{\hat l_{i,j}},{c_{i,j}} = 0} )}$ and $\scalemath{0.9}{p( {\bar \mu _{i,j}^{\row,(\ell )}|{\hat l_{i,j}},{c_{i,j}} = 1})}$ are possible ($12$ in total). The term $\scalemath{0.9}{p( {\bar \mu _{i,j}^{r,(\ell )}=-1|{\hat l_{i,j}=-1},{c_{i,j}} = 0} )}$ is the probability of error at the BDD output given that the channel output is also in error. One can check that this is exactly the definition of $\scalemath{0.9}{p( {\bar \mu _{i,j}^{\row,(\ell )}=1|{\hat l_{i,j}=1},{c_{i,j}} = 1} )}$
 as the values for $\bar \mu _{i,j}^{\row,(\ell )}$ and $\hat l_{i,j}$ are changed from $-1$ to $1$ and the value for $c_{i,j}$ is changed from $0$ to $1$. Therefore, $\scalemath{0.9}{p( {\bar \mu _{i,j}^{\row,(\ell )}=-1|{\hat l_{i,j}=-1},{c_{i,j}} = 0} )}=\scalemath{0.9}{p( {\bar \mu _{i,j}^{\row,(\ell )}=1|{\hat l_{i,j}=1},{c_{i,j}} = 1} )}$. Similarly, the following relations also hold: \\
$\scalemath{0.9}{p( {\bar \mu _{i,j}^{\row,(\ell )}=1|{\hat l_{i,j}=1},{c_{i,j}} = 0} )}=\scalemath{0.9}{ p( {\bar \mu _{i,j}^{\row,(\ell )}=-1|{\hat l_{i,j}=-1},{c_{i,j}} = 1} )}$, $\scalemath{0.9}{ p( {\bar \mu _{i,j}^{\row,(\ell )}=1|{\hat l_{i,j}=-1},{c_{i,j}} = 0} )}=\scalemath{0.9}{ p( {\bar \mu _{i,j}^{r,(\ell )}=-1|{\hat l_{i,j}=1},{c_{i,j}} = 1} )}$, $\scalemath{0.9}{p( {\bar \mu _{i,j}^{\row,(\ell )}=-1|{\hat l_{i,j}=1},{c_{i,j}} = 0} )}=\scalemath{0.9}{p( {\bar \mu _{i,j}^{r,(\ell )}=1|{\hat l_{i,j}=-1},{c_{i,j}} = 1} )}$, $\scalemath{0.9}{p( {\bar \mu _{i,j}^{\row,(\ell )}=0|{\hat l_{i,j}=-1},{c_{i,j}} = 0} )}=\scalemath{0.9}{p( {\bar \mu _{i,j}^{\row,(\ell )}=0|{\hat l_{i,j}=1},{c_{i,j}} = 1} )}$, and $\scalemath{0.85}{p( {\bar \mu _{i,j}^{\row,(\ell )}=0|{\hat l_{i,j}=1},{c_{i,j}} = 0} )}=\scalemath{0.85}{p( {\bar \mu _{i,j}^{\row,(\ell )}=0|{\hat l_{i,j}=-1},{c_{i,j}} = 1})}$.

We assume all-zero codeword transmission. Such assumption yields the computation of six different terms corresponding to $\scalemath{0.9}{p( {\bar \mu _{i,j}^{\row,(\ell )}|{\hat l_{i,j}},{c_{i,j}} = 0})}$.

\SHm{Let $\mep$ be  the input error probability to the row-type CNs for the BDD stage and  denote by $f^{\Pue}(\mep)$} the probability that a \emph{randomly selected bit} in the component code's codeword is decoded incorrectly when it was \emph{initially in error}. \SHm{The notion of \emph{randomly selected bit} is motivated due to the fact that we analyze the ensemble given in Fig.~\ref{PCgraph}(c), where the connection between VN and CN is randomly built.  Also, we use the notion of \emph{initially in error} as we analyze the extrinsic message passing where in each iteration the input corresponding to $c_{i,j}$ is substituted by channel input $r_{i,j}$ (see \eqref{eq:BDD_VN_extrinsic}).}
Furthermore, we denote by $\Pue\left( i \right)$ the probability that a \emph{randomly selected bit} in the component code's codeword is decoded incorrectly when it was \emph{initially} in error and there are $i$ errors in the other $n-1$ positions. \AG{$\Pue\left( i \right)$ was derived   in \cite[Eq. (5)]{sheikhTCOM19}  for iBDD-SR and the same expression holds for iBDD-CR. 
$f^{\Pue}(\mep)$ is then obtained based on the union of (independent) events for  $i \in \{0,\cdots,n-1\}$ random errors  as}
\begin{align}\label{pe} 
\scalemath{0.86}{ f^{\Pue}(\mep) \buildrel \Delta \over =  p(\bar \mu _{i,j}^{\row,(\ell )}=-1|{\hat l_{i,j}}=-1, c_{i,j}=0) =  \sum\limits_{i = 0}^{n - 1}   b^{n}_i(x) {\Pue\left( i \right)}}.
\end{align} 
where $b^{n}_i(x)\buildrel \Delta \over  = {{n-1}\choose{i}} {\mep^i}{\left( {1 - \mep} \right)^{n - i - 1}}$.

We denote by $f^{\Puc}(\mep)$ and $f^{\Puep}(\mep)$ the probability that a randomly selected bit in the component code's codeword is decoded correctly and erased,\footnote{Note the $\bar \mu _{i,j}^{\row,(\ell )}=0$ corresponds to the erasure output of BDD.} respectively, when it was initially in error. Furthermore, we denote by $\Puc\left( i \right)$ and $P^\epsilon(i)$ the probability that a randomly selected bit in the
component code's codeword is decoded correctly and erased, respectively, when it was initially in error and there are $i$  errors in the other $n-1$ positions. $\Puc\left( i \right)$ is \AG{given} in \cite[Eq. (9)]{sheikhTCOM19} and $P^\epsilon(i)=1-\Pue\left( i \right)-\Puc\left( i \right)$. Similar to the derivation of \eqref{pe},  $f^{\Puc}(\mep)$ and $f^{\Puep}(\mep)$ \AG{are obtained as}
\begin{align}\label{pc} 
\scalemath{0.86}{ f^{\Puc}(\mep) \buildrel \Delta \over  =  p(\bar \mu _{i,j}^{\row,(\ell )}=1|{\hat l_{i,j}}=-1, c_{i,j}=0)   =  \sum\limits_{i = 0}^{n - 1}  b^{n}_i(x) {\Puc\left( i \right)}},
\end{align} 
\begin{align}\label{p_ep} 
\scalemath{0.86}{f^{\Puep}(\mep) \buildrel \Delta \over  =  p(\bar \mu _{i,j}^{\row,(\ell )}=0|{\hat l_{i,j}}=-1,c_{i,j}=0)  =  \sum\limits_{i = 0}^{n - 1}  b^{n}_i(x)  {\Puep\left( i \right)}}.
\end{align}
Furthermore, we denote by $f^{\Que}(\mep)$, $f^{\Quc}(\mep)$, and $f^{\Quep}(\mep)$ 
the probability that a randomly selected bit in the component code's codeword is decoded incorrectly, correctly, and erased, respectively, when the bit was initially correct. Also, let us denote by $\Que\left( i \right)$, $\Quc\left( i \right)$, and $Q^\epsilon(i)$ 
the probability that a randomly selected bit in the component code's codeword is decoded incorrectly, correctly, and erased, respectively, when the bit was initially correct and there are $i$ errors in the remaining $n-1$ positions. $\Quc\left( i \right)$ is \AG{given in \cite[Eq. (6)]{sheikhTCOM19}, $\Que\left( i \right)$   in \cite[Eq. (10)]{sheikhTCOM19},} and $\Quep\left( i \right)=1-\Que\left( i \right)-\Quc\left( i \right)$. Following the same steps \AG{as for the  derivation of \eqref{pe}, we get}
\begin{align}\label{qe} 
\scalemath{0.86}{f^{\Que}(\mep) \buildrel \Delta \over  =  p(\bar \mu _{i,j}^{\row,(\ell )}=-1|{\hat l_{i,j}}=1,c_{i,j}=0)   =  \sum\limits_{i = 0}^{n - 1}   b^{n}_i(x) {\Que\left( i \right)}},
\end{align}
\begin{align}\label{qc} 
\scalemath{0.86}{f^{\Quc}(\mep)  \buildrel \Delta \over  =  p(\bar \mu _{i,j}^{\row,(\ell )}=1|{\hat l_{i,j}}=1,c_{i,j}=0)   =  \sum\limits_{i = 0}^{n - 1}   b^{n}_i(x)  {\Quc\left( i \right)}},
\end{align}
\begin{align}\label{q_ep} 
\scalemath{0.86}{f^{\Quep}(\mep)  \buildrel \Delta \over  =  p(\bar \mu _{i,j}^{\row,(\ell )}=0|{\hat l_{i,j}}=1,c_{i,j}=0)  =  \sum\limits_{i = 0}^{n - 1}   b^{n}_i(x)  {\Quep\left( i \right)}}.
\end{align}
Recalling \eqref{mutild_r} and the discussion given in the begining of this appendix, by employing \eqref{pe}--\eqref{q_ep} and $\mep^{\col,(\ell-1)}$ as the input error probability of row-type CN at iteration $\ell$, the values of $\tilde \mu _{i,j}^{\row,(\ell )}$ given in Table~\ref{Tabcomp} \AG{are obtained}. 

   
Now we focus on computing $\mep^{\row,(\ell)}$. \AG{Assuming transmission of the all-zero codeword, using \eqref{LLRoutv1}, and applying Bayes' rule}, the probability of error at the output of row decoder is obtained as 
\begin{align}\label{p_bays} 
\mep^{\row,(\ell)} &= p(\opt_{i,j}^{\mathsf r, (\ell)}<0) \nonumber \\ &= \sum\limits_{\mathclap{\substack{
			{\bar \mu _{i,j}^{\row,(\ell )}\in \{ 0, \pm 1\} }\\
			{{{\hat L}_{i,j}}\in \{  \pm 1\} }}}} {p(\opt_{i,j}^{\mathsf r, (\ell)} < 0|\bar \mu _{i,j}^{\row,(\ell )},{{\hat l}_{i,j}})} p(\bar \mu _{i,j}^{\row,(\ell )}|{\hat l_{i,j}})p({\hat l_{i,j}})
\end{align}
\AG{where  $p(\bar \mu _{i,j}^{\row,(\ell )}|{\hat l_{i,j}})$ is given in \eqref{pe}--\eqref{q_ep}}. In the following, we compute ${p(\opt_{i,j}^{\mathsf r, (\ell)} < 0|\bar \mu _{i,j}^{\row,(\ell )},{{\hat l}_{i,j}})}$. In particular, let us first calculate the term ${p(\opt_{i,j}^{\mathsf r, (\ell)}< 0|\bar \mu _{i,j}^{\row,(\ell )}=-1,{{\hat l}_{i,j}=-1})}$ in \eqref{p_bays}. This probability can be written as
\begin{align}\label{p_baysd1} 
& \nonumber {p\left(\opt_{i,j}^{\mathsf r, (\ell)} < 0\lvert\bar \mu _{i,j}^{\row,(\ell )}=-1,{{\hat l}_{i,j}=-1}\right)}\myeqa \\ \nonumber & {p\left( l_{i,j} < -\ln \left( \frac{f^{\Pue}(\mep^{\col,(\ell-1)})}{f^{\Quc}(\mep^{\col,(\ell-1)})}\right)\bigg\lvert{\bar\mu _{i,j}^{\row,(\ell )}=-1,{\hat l}_{i,j}}=-1\right)}\myeqb  \\ \nonumber & {p\left( l_{i,j} < -\ln \left( \frac{f^{\Pue}(\mep^{\col,(\ell-1)})}{f^{\Quc}(\mep^{\col,(\ell-1)})}\right)\bigg\lvert{{\hat l}_{i,j}}=-1\right)}\myeqc \\ \nonumber & \frac{p\left(l_{i,j}<-\ln \left( \frac{f^{\Pue}(\mep^{\col,(\ell-1)})}{f^{\Quc}(\mep^{\col,(\ell-1)})}\right),l_{i,j}<0\right)}{p(l_{i,j}<0)} \myeqd \\ \nonumber & \frac{p\left(l_{i,j}<\text{min}\left(-\ln \left( \frac{f^{\Pue}(\mep^{\col,(\ell-1)})}{f^{\Quc}(\mep^{\col,(\ell-1)})}\right),0\right)\right)}{p(l_{i,j}<0)} \myeqe \\ & \frac{{\Q\left( { \frac{\sigma}{2}\text{min}\left(\ln \left( \frac{f^{\Pue}(\mep^{\col,(\ell-1)})}{f^{\Quc}(\mep^{\col,(\ell-1)})}\right),0\right)+\frac{1}{\sigma}} \right) }}{p(l_{i,j}<0)}
\end{align}
where $(a)$ follows from \eqref{LLRoutv1} and  Table~\ref{Tabcomp}, $(b)$ follows from the Markov chain $l_{i,j} \to \hat l_{i,j} \to \bar \mu _{i,j}^{\row,(\ell )}$ (see Sec.~\ref{sec:DE_BIAWGN}), $(c)$ follows from the definition of conditional probability and the fact that $p(\hat l_{i,j}=-1)=p(l_{i,j}<0)$, $(d)$ follows by intersecting the events $\Big\{l_{i,j}< -\ln \left( \frac{f^{\Pue}(\mep^{\col,(\ell-1)})}{f^{\Quc}(\mep^{\col,(\ell-1)})}\right)\Big\}$ and $\{l_{i,j}<0\}$, and $(e)$ follows by recalling that $l_{i,j}\sim\mathcal{N}(2/\sigma^2,4/\sigma^2)$ if $c_{i,j}=0$ and employing the $\Q(\cdot)$ function. 

With the same approach above, one can compute the following probabilities
\begin{align}\label{p_baysd2} 
& \nonumber {p\left(\opt_{i,j}^{\mathsf r, (\ell)} < 0|\bar \mu _{i,j}^{\row,(\ell )}=1,{{\hat l}_{i,j}=-1}\right)}= \\ &  ~~~~~~~~~\frac{{\Q\left( { \frac{\sigma}{2}\text{min}\left(\ln \left( \frac{f^{\Puc}(\mep^{\col,(\ell-1)})}{f^{\Que}(\mep^{\col,(\ell-1)})}\right),0\right)+\frac{1}{\sigma}} \right) }}{p(l_{i,j}<0)},
\end{align}
\begin{align}\label{p_baysd3} 
& \nonumber {p\left(\opt_{i,j}^{\mathsf r, (\ell)} < 0|\bar \mu _{i,j}^{\row,(\ell )}=0,{{\hat l}_{i,j}=-1}\right)}= \\ &  ~~~~~~~~~\frac{{\Q\left( { \frac{\sigma}{2}\text{min}\left(\ln \left(\frac{f^{\Puep}(\mep^{\col,(\ell-1)})}{f^{\Quep}(\mep^{\col,(\ell-1)})}\right),0\right)+\frac{1}{\sigma}} \right) }}{p(l_{i,j}<0)}.
\end{align}
The probability ${p\left(\opt_{i,j}^{\mathsf r, (\ell)} < 0|\bar \mu _{i,j}^{\row,(\ell )}=-1,{{\hat l}_{i,j}=1}\right)}$ is 
\begin{align}\label{p_baysd4} 
& \nonumber {p\left(\opt_{i,j}^{\mathsf r, (\ell)} < 0\lvert\bar \mu _{i,j}^{\row,(\ell )}=-1,{{\hat l}_{i,j}=1}\right)}\myeqa \\ \nonumber & {p\left( l_{i,j} < -\ln \left( \frac{f^{\Que}(\mep^{\col,(\ell-1)})}{f^{\Puc}(\mep^{\col,(\ell-1)})}\right)\bigg\lvert\bar \mu _{i,j}^{\row,(\ell )}=-1,{{\hat l}_{i,j}}=1\right)}\myeqb \\ \nonumber & {p\left( l_{i,j} < -\ln \left( \frac{f^{\Que}(\mep^{\col,(\ell-1)})}{f^{\Puc}(\mep^{\col,(\ell-1)})}\right)\bigg\lvert{{\hat l}_{i,j}}=1\right)}\myeqc \\ & \frac{p\left(l_{i,j}<-\ln \left( \frac{f^{\Que}(\mep^{\col,(\ell-1)})}{f^{\Puc}(\mep^{\col,(\ell-1)})}\right), l_{i,j}>0 \right)}{p(l_{i,j}>0)} \myeqd \\ & \frac{p\left(0 < l_{i,j}<-\ln \left( \frac{f^{\Que}(\mep^{\col,(\ell-1)})}{f^{\Puc}(\mep^{\col,(\ell-1)})}\right) \right)}{p(l_{i,j}>0)},
\end{align}
where $(a)$ follows from \eqref{LLRoutv1} and Table~\ref{Tabcomp}, $(b)$ follows from the Markov chain $l_{i,j} \to \hat l_{i,j} \to \bar \mu _{i,j}^{\row,(\ell )}$, $(c)$ follows from the definition of conditional probability and the fact that $p(\hat l_{i,j}=1)=p(l_{i,j}>0)$, and $(d)$ follows by intersecting the events $\Big\{l_{i,j}< -\ln \left( \frac{f^{\Que}(\mep^{\col,(\ell-1)})}{f^{\Puc}(\mep^{\col,(\ell-1)})}\right)\Big\}$ and $\{l_{i,j}>0\}$. Note that we assumed that $\ln \left(\frac{f^{\Que}(\mep^{\col,(\ell-1)})}{f^{\Puc}(\mep^{\col,(\ell-1)})}\right)<0$, as for $\ln \left(\frac{f^{\Que}(\mep^{\col,(\ell-1)})}{f^{\Puc}(\mep^{\col,(\ell-1)})}\right)>0$ the intersection between the events $\Big\{l_{i,j}< -\ln \left( \frac{f^{\Que}(\mep^{\col,(\ell-1)})}{f^{\Puc}(\mep^{\col,(\ell-1)})}\right)\Big\}$ and $\{l_{i,j}>0\}$ is null and \eqref{p_baysd4} boils down to zero. By recalling the distribution of $l_{i,j}$ and employing $\Q(\cdot)$, we have
\begin{align}\label{p_baysd44}
& \nonumber {p\left(\opt_{i,j}^{\mathsf r, (\ell)} < 0|\bar \mu _{i,j}^{\row,(\ell )}=-1,{{\hat l}_{i,j}=1}\right)}= \\ &  
\scalemath{0.85}{\frac{{\Q\left( { \frac{\sigma}{2}\ln \left(\frac{f^{\Que}(\mep^{\col,(\ell-1)})}{f^{\Puc}(\mep^{\col,(\ell-1)})}\right)+\frac{1}{\sigma}} \right)}-\Q\left(\frac{1}{\sigma}\right)}{p(l_{i,j}>0)}\cdot \bar{\mathbbm{U}} \left( \ln \left(\frac{f^{\Que}(\mep^{\col,(\ell-1)})}{f^{\Puc}(\mep^{\col,(\ell-1)})}\right) \right)}
\end{align}
Similarly, one can compute the following probabilities
\begin{align}\label{p_baysd5}
& \nonumber {p\left(\opt_{i,j}^{\mathsf r, (\ell)} < 0|\bar \mu _{i,j}^{\row,(\ell )}=1,{{\hat l}_{i,j}=1}\right)}= \\ & \scalemath{0.83}{\frac{{\Q\left( { \frac{\sigma}{2}\ln \left(\frac{f^{\Quc}(\mep^{\col,(\ell-1)})}{f^{\Pue}(\mep^{\col,(\ell-1)})}\right)+\frac{1}{\sigma}} \right)}-\Q\left(\frac{1}{\sigma}\right)}{p(l_{i,j}>0)}\cdot \bar{\mathbbm{U}} \left( \ln \left(\frac{f^{\Quc}(\mep^{\col,(\ell-1)})}{f^{\Pue}(\mep^{\col,(\ell-1)})}\right) \right)}
\end{align}
\begin{align}\label{p_baysd6}
& \nonumber {p\left(\opt_{i,j}^{\mathsf r, (\ell)} < 0|\bar \mu _{i,j}^{r,(\ell )}=0,{{\hat l}_{i,j}=1}\right)}= \\ &  \scalemath{0.83}{\frac{{\Q\left( { \frac{\sigma}{2}\ln \left(\frac{f^{\Quep}(\mep^{\col,(\ell-1)})}{f^{\Puep}(\mep^{\col,(\ell-1)})}\right)+\frac{1}{\sigma}} \right)}-\Q\left(\frac{1}{\sigma}\right)}{p(l_{i,j}>0)}\cdot \bar{\mathbbm{U}} \left( \ln \left(\frac{f^{\Quep}(\mep^{\col,(\ell-1)})}{f^{\Puep}(\mep^{\col,(\ell-1)})}\right) \right)}
\end{align}

By substituting \eqref{pe}--\eqref{q_ep} and \eqref{p_baysd1}--\eqref{p_baysd6} into \eqref{p_bays}, the closed-form expression for $\mep^{\row,(\ell)}$ in \eqref{xrl} is obtained. We remark that in this substitution, $p({\hat l_{i,j}})$ in \eqref{p_bays} cancels out with the denominator of \eqref{p_baysd1}--\eqref{p_baysd6}.  


By substituting $\mep^{\col,(\ell-1)}$, $\bar \mu _{i,j}^{\row,(\ell )}$, and $\opt_{i,j}^{\mathsf r, (\ell)}$ with $\mep^{\row,(\ell)}$, $\bar \mu _{i,j}^{\col,(\ell )}$, and $\opt_{i,j}^{\mathsf c, (\ell)}$, and following the same derivation steps as above, Table~\ref{Tabcompc} and \eqref{xcl} are obtained. This concludes the proof.

\section{Proof of \GL{Proposition} 2} \label{AB} 
\AG{By considering channel adapters, we can  assume the transmission of the all-zero codeword  in the DE analysis. With this assumption  and employing \eqref{LLRMaxlog_approx} and \eqref{LLRMaxlog_approxllr}, the PDF of the symmetrized LLRs is}
\begin{align} \label{LLRMaxlog_approx1}
p\left( {\bar l_{i,j}|b_{i,j} = 0} \right) = \sum\limits_{j = 0}^{\frac{M}{2} - 1} {{w_j}} {\G}(\bar l;{\mu _j},\sigma _j^2),
\end{align}
where $w_j$, $\mu _j$, and $\sigma _j^2$ are given in Sec.~\ref{QAM_BICM}. 
The \AG{main difference between the DE analysis for BICM and for the bi-AWGN channel  in Appendix~\ref{APP1} is that  the PDF of the LLRs in \eqref{LLRMaxlog_approx1} should be used in the analysis. By employing \eqref{LLRMaxlog_approx1} in \eqref{p_baysd1}--\eqref{p_baysd6} the different terms for \eqref{p_bays} are obtained and the expressions in \eqref{QAMxrl} and \eqref{QAMxcl} result for $\mep^{\row,(\ell)}$ and $\mep^{\col,(\ell)}$, respectively. We remark that to run DE  the error probability of the VNs should be initialized to $p_\mathsf{ch}$}. For the BICM channel $p_\mathsf{ch}$ can be computed by integrating the tail of the LLR distribution \eqref{LLRMaxlog_approx1}, yielding
\begin{align} \label{pch_computation}
p_\mathsf{ch} = p(\bar l_{i,j}<0|b_{i,j} = 0)=\sum\limits_{j = 0}^{\frac{M}{2} - 1} {{w_j}}\cdot\Q\left(\frac{\mu_j}{\sigma_j}\right).
\end{align}  
This concludes the proof.

\end{document}